\documentclass[journal]{IEEEtran}
\IEEEoverridecommandlockouts
\usepackage{cite}
\usepackage[hidelinks]{hyperref}
\usepackage{amsmath,amssymb,amsfonts}
\usepackage{algorithmic} 
\usepackage{graphicx}
\usepackage{textcomp}
\usepackage{xcolor}
\usepackage{CJK}
\usepackage{algorithm}
\usepackage{pifont}
\usepackage{amsthm}
\usepackage{subfig}
\usepackage{float}  %
\usepackage{makecell}
\usepackage{booktabs} 

\usepackage{algorithmic}
\usepackage{amsmath}
\usepackage{amssymb}
\usepackage{psfrag}
\usepackage{graphicx}
\usepackage{epstopdf}
\usepackage{multirow}
\usepackage{stfloats}
\usepackage{color}
\usepackage[normalem]{ulem}
\usepackage{arydshln}
\usepackage{enumerate}
\usepackage{bbm}
\usepackage{verbatim}

\begin{document}
\title{User Association and Coordinated Beamforming in Cognitive Aerial-Terrestrial Networks: \\A Safe Reinforcement Learning Approach\\
}

\author{
Zizhen Zhou, Jungang Ge, and Ying-Chang Liang, \IEEEmembership{Fellow, IEEE}


\thanks{
Part of this work was presented in IEEE GLOBECOM 2024 \cite{zhou2024dynamic}.

Z. Zhou is with the National Key Laboratory of Wireless Communications, University of Electronic Science and Technology of China (UESTC), Chengdu 611731, China (e-mail: zhouzizhen@std.uestc.edu.cn). 

J. Ge is with the Department of Mobile Communications and Terminal Research, China Telecom Research Institute, Guangzhou 510000, China (e-mail: gejg1@chinatelecom.cn).

Y.-C. Liang is with the Center for Intelligent Networking and Communications (CINC), University of Electronic Science and Technology of China (UESTC), Chengdu 611731, China (e-mail: liangyc@ieee.org).
}
}

\maketitle

\begin{abstract}
Cognitive aerial-terrestrial networks (CATNs) facilitate the sharing of spectrum resources between aerial and terrestrial networks, presenting a promising solution to the spectrum scarcity challenges posed by thriving aerial networks.
However, aerial users (AUs), such as airplanes and flying cars, which demand high-quality downlink communication, experience significant interference from numerous terrestrial base stations (BSs).
To alleviate such interference, in this paper, we investigate a user association and coordinated beamforming (CBF) problem in CATN, where the aerial network serves as the primary network sharing its spectrum with the terrestrial network.
Specifically, we maximize the sum rate of the secondary terrestrial users (TUs) while satisfying the interference temperature constraints of the AUs.
Traditional iterative optimization schemes are impractical for solving this problem due to their high computational complexity and information exchange overhead.
Although deep reinforcement learning (DRL) based schemes can address these challenges, their performance is sensitive to the weights of the weighted penalty terms for violating constraints in the reward function.
Motivated by these issues, we propose a safe DRL-based user association and CBF scheme for CATN, eliminating the need for training multiple times to find the optimal penalty weight before actual deployment.
Specifically, the CATN is modeled as a networked constrained partially observable Markov game.
Each TU acts as an agent to choose its associated BS, and each BS acts as an agent to decide its beamforming vectors, aiming to maximize the reward while satisfying the safety constraints introduced by the interference constraints of the AUs.
By exploiting a safe DRL algorithm, the proposed scheme incurs lower deployment expenses than the penalty-based DRL schemes since only one training is required before actual deployment.
Simulation results show that the proposed scheme can achieve a higher sum rate of TUs than a two-stage optimization scheme while the average received interference power of the AUs is generally below the threshold.

\end{abstract}


\section{Introduction}
\label{sec_introduction}
%
%

Aerial networks are experiencing rapid growth, leveraging the flexible deployment advantages of various aircraft to facilitate a wide range of applications, including the Internet of Things (IoT), transportation, logistics, tourism, agriculture, healthcare, rescue, and disaster monitoring \cite{ jiang20236g}. 
Aircrafts, such as unmanned aerial vehicles (UAVs), airplanes, or flying cars, can act as the aerial users (AUs) communicating with base stations (BSs).
For example, air-to-ground (ATG) networks can provide in-flight communications to passengers by deploying dedicated BSs on the ground \cite{baltaci2021survey, mozaffari2021toward}.
%
%
Besides, flying cars, e.g., electric vertical takeoff and landing (eVTOL) aircrafts, enable urban air mobility and provide infotainment for passengers \cite{pan2021flying, zaid2023evtol}. 
These emerging services bring the increasing downlink communication needs, which pose challenges for effective spectrum utilization \cite{baltaci2021survey, zaid2023evtol}.
%
%
To improve spectrum utilization efficiency, cognitive spectrum sharing offers a promising solution by enabling secondary networks to access the spectrum of the primary network if the received interference power of the primary user remains below a certain threshold, a.k.a., the interference temperature limit \cite{liang2011cognitive}.
Nonetheless, when terrestrial cellular networks and aerial networks use the same spectrum for transmission, the AUs face significant interference from numerous terrestrial BSs due to the line-of-sight (LoS)-dominant channels \cite{baltaci2021survey, tadayon2016inflight, zeng2018cellular, 3GPP38876}. 

%


To satisfy the interference temperature constraints of the AUs, BSs can perform coordinated beamforming (CBF), where the beamformers of the BSs are optimized by some iterative methods, e.g., the iterative weighted minimum mean square error (WMMSE) algorithm \cite{shi2011iteratively} and the fractional programming (FP) algorithms \cite{shen2018fractional1}.
%
%
%
%
To further mitigate interference, the user association (UA) is optimized to enable users to associate with the BS that has a strong channel, thereby preventing this channel from becoming a strong interference link \cite{sanjabi2014optimal}.
%
A common UA method is based on received signal strength (RSS), where the user equipment (UE) tends to access the BS with the strongest RSS \cite{mahmoud2010performance, wang2018handover}.
Unfortunately, numerous UEs may attempt to access the same BS, leading to access congestion and reduced throughput. 
To address this issue, the WMMSE algorithm is modified to jointly optimize UA and beamforming vectors by collecting the real-time global channel state information (CSI) for centralized decision-making \cite{sanjabi2014optimal}.
Nonetheless, this modified WMMSE algorithm still suffers from high computational complexity and introduces excessive handover.
As a solution, by decoupling UA and beamforming, a two-stage joint UA and WMMSE beamforming algorithm is proposed in \cite{shen2014distributed}, achieving almost the same sum-rate performance as that in \cite{sanjabi2014optimal}.
In addition, the minimum weighted signal-to-interference-plus-noise ratio (SINR) is maximized in \cite{xie2017sinr}.
In \cite{dong2017joint}, with only the statistical CSI, the ergodic sum rate of all UEs is maximized subject to the SINR constraint of each UE.
%
However, these traditional iterative optimization algorithms require real-time global CSI (except for \cite{dong2017joint}) and exhibit high computational complexity, making them difficult to apply in practice, particularly with fast-moving AUs \cite{sanjabi2014optimal, shen2014distributed, xie2017sinr, dong2017joint}.

Recently, deep reinforcement learning (DRL) has been utilized to solve UA problems \cite{wang2018handover, cao2020deep, zhao2019deep} and beamforming problems \cite{liu2022deep, ge2024deep}, demonstrating advantages in computational complexity and information acquisition.
%
In \cite{cao2020deep}, each UE acts as a deep Q-network (DQN) agent and only needs to know the number of access users of all BSs in the previous time slot, without requiring real-time global RSS information.
In \cite{zhao2019deep}, a DRL-based joint UA and channel allocation scheme is proposed, where each UE acts as a dueling double DQN (D3QN) agent and only requires knowing whether the quality of service (QoS) requirements of other UEs are met.
In \cite{ge2024deep}, a DRL-based distributed dynamic CBF scheme is proposed for the massive MIMO cellular network, where each BS acts as an agent to serve multiple single-antenna users in its cell without requiring real-time global CSI.
When facing constrained CBF problems in cognitive radio networks, DRL algorithms can also be utilized, where the reward function contains the weighted penalty terms for the violation of the interference temperature constraints \cite{zhao2019deep}.
However, the performance of these penalty-based DRL algorithms is sensitive to the penalty weight in some applications \cite{achiam2017constrained, li2019constrained}.
Specifically, when the penalty weight is large, the algorithm is too conservative to achieve a high reward. 
When the penalty weight is small, the algorithm is too bold to effectively satisfy the constraint.
Therefore, it is necessary to train multiple times to find the optimal penalty weight before actual deployment, which incurs large expenses.
Besides, the reward function with a penalty term complicates the value function, which makes it difficult for the critic network to converge \cite{wu2024real}.
%
%
%
To ensure safety when deploying DRL in real-world applications, the constrained Markov decision process (CMDP) with cost function sets has been studied \cite{gu2024review}.
To solve the CMDP problems, safe DRL algorithms have been developed to maximize the expected cumulative discounted reward while satisfying safety constraints, namely, the expected cumulative discounted costs are lower than the corresponding cost limits \cite{gu2024review}.
Without finding the optimal penalty weights described above, the safe DRL-based approaches exhibit lower deployment expenses.

Motivated by the above considerations, we propose a safe DRL-based distributed dynamic UA and CBF (DDUACBF) framework for the cognitive aerial-terrestrial networks (CATNs).
In CATN, the aerial network acts as a primary network to share its spectrum with the secondary terrestrial network if the interference from the terrestrial network can be tolerated.
We formulate a problem to maximize the sum rate of the terrestrial users (TUs) by optimizing the user association between TUs and BSs and the beamforming vectors of the terrestrial BSs under the interference temperature constraints of the AUs, i.e., the primary users.
Then, we model this problem as a networked constrained partially observable Markov game (NCPOMG) \cite{zhang2021decentralized, feriani2021single, gu2023safe}, which contains two types of agents, i.e., each TU is a UA agent to decide which BS to associate with, and each BS is a beamforming agent to decide the beamforming vectors for its associated TUs.
To alleviate the difficulties caused by non-stationarity and partial observability, we design observation of agents by enabling information exchange between agents.
Besides, the dimension of the action space for BS agents is reduced by exploiting a known solution structure derived from the traditional optimization algorithms \cite{ge2024deep}.
In particular, the shared cost functions are introduced for all BS agents, which are linear functions of the received interference power of the AUs.
Finally, we propose a safe DRL-based scheme to solve the NCPOMG, where a safe DRL algorithm is applied to the BS agents to maximize the reward while satisfying the safety constraints, i.e., the expected cumulative discounted received interference power of the AUs are below the thresholds.
The main contributions of the paper are summarized as follows:
\begin{itemize}
\item We study the user association and beamforming problem in the cognitive aerial-terrestrial network, where under the interference temperature constraints of the AUs, the beamforming vectors of the terrestrial BSs and the user association are optimized to maximize the sum rate of the TUs.
\item We develop a multi-agent DRL-based user association and beamforming framework for CATN.
In particular, the studied problem is modeled as an NCPOMG, where each TU is a UA agent and each BS is a beamforming agent.
To the best of our knowledge, DRL-based user association and beamforming is studied in the non-joint transmission scenario for the first time.
\item 
A safe DRL algorithm is applied to the BS agents, which maximizes the reward while satisfying the safety constraints related to the interference temperature constraints of the AUs.
This approach significantly reduces deployment expenses since it avoids training multiple times to find the optimal penalty weight before the actual deployment, which is required by the penalty-based DRL method.
\item Simulation results show that the proposed safe DRL-based scheme can achieve a higher sum rate of the TUs than a two-stage optimization scheme while the safety constraints are well satisfied.
Moreover, the proposed scheme exhibits lower computational complexity than the iterative optimization-based schemes and does not require real-time global CSI.

\end{itemize}

The rest of this paper is organized as follows:
Section \ref{secRelatedWork} presents the related work.
In Section \ref{secSystemModel}, the system model of CATN considered in this study is introduced.
In Section \ref{secProblem}, a user association and beamforming problem in CATN is formulated and a two-stage optimization-based scheme is introduced.
In Section \ref{secproposedScheme}, we first introduce the NCPOMG and safe reinforcement learning, and then propose a safe DRL-based DDUACBF framework for the CATN.
Section \ref{sec_Simulation_Results} shows the simulation results.
Finally, Section \ref{sec_Conclusions} concludes this paper.


Notations used in this paper are listed as follows.
The lowercase, bold lowercase, and bold uppercase, i.e., $a$, ${\bf{a}}$, and ${\bf{A}}$ are scalar, vector, and matrix, respectively.
$\mathbb{C}^{a \times b}$ denotes the space of $a \times b$ complex-valued matrices.
${\bf{I}}$ denotes an identity matrix.
$| \cdot |$ denotes the absolute value. 
${{\left\| {\bf{a}} \right\|}_2}$ denotes the $\ell_2$ norm of vector ${\bf{a}}$.
$(\cdot )^T$ and $(\cdot )^H$ denote transpose and conjugate transpose, respectively.
${\mathbb{E}}\{\cdot\}$ denotes the average operation.
${\mathcal {CN}} (\mu, \sigma^2)$ denotes the complex Gaussian distribution with mean $\mu$ and variance $\sigma^2$.
${\rm{clip}}(x,a,b)$ denotes a function that clips $x$ to the interval $[a,b]$.
${\bf{a}} \odot {\bf{b}}$ denotes the element-wise product, which means multiplying the corresponding elements of vector ${\bf{a}}$ and vector ${\bf{b}}$.

\section{Related Work}
\label{secRelatedWork}
Interference management for the AUs, e.g., UAVs, has been extensively studied \cite{zeng2018cellular, mei2021aerial, mei2019cooperative}.
In \cite{zeng2018cellular}, the three-dimensional maximal-ratio transmission beamforming is adopted by BSs to serve UAVs and ground users on the shared channel.
The work in \cite{mei2021aerial} reviews traditional aerial-ground interference mitigation solutions, including inter-cell cooperation, dynamic frequency reuse, coordinated multipoint (CoMP), non-orthogonal multiple access (NOMA), beamforming, along with their respective drawbacks.
Then, solutions utilizing the sensing ability of UAVs and idle BSs in the network are proposed.
In \cite{mei2019cooperative}, downlink cooperative beamforming with interference transmission and cancellation is proposed to mitigate the interference to the UAV caused by the co-channel terrestrial transmissions.

The DRL-based downlink interference management in aerial-terrestrial networks has been explored, where the cellular-connected UAVs and terrestrial UEs coexist \cite{li2023radio, burhanuddin2023inter}.
In \cite{li2023radio}, to minimize the ergodic outage duration of the UAV, resource block allocation and downlink beamforming for UAV are determined by the D3QN agent in large-scale and an agent in small-scale respectively.
In \cite{burhanuddin2023inter}, to maximize the throughput of terrestrial UEs under the transmission rate threshold for UAVs and terrestrial UEs, the number of muting cells and the slice time allocation are determined by a DQN agent.

The joint UA and power control based on multi-agent DRL (MADRL) has been extensively studied \cite{guo2020joint, naderializadeh2021resource, alwarafy2022hierarchical, yang2022distributed}.
To address the challenge of environmental non-stationary posed by independent agents in \cite{wang2018handover}, the handover and power allocation problem is modeled as a fully cooperative multi-agent task in \cite{guo2020joint}, which is solved by a multi-agent proximal policy optimization (PPO) algorithm based on the centralized training with decentralized execution (CTDE) framework where each UE acts as an agent.
In \cite{naderializadeh2021resource}, a MADRL framework with scalable observation and action spaces is developed, where each access point (AP) acts as an agent.
In \cite{alwarafy2022hierarchical}, a hierarchical MADRL-based framework is proposed, where an edge server agent determines the UA, while multiple AP agents manage power control.
In \cite{yang2022distributed}, device association, spectrum allocation, and power allocation in heterogeneous networks are optimized, where each BS acts as a D3QN agent.

A joint UA and beamforming scheme based on MADRL is proposed in \cite{yu2023distributed}, which adopts a two-timescale framework, i.e., UA and beamforming are determined on large and small time scales respectively.
This scheme is tailored for the joint transmission scenario, where one UE can be served by multiple APs.
Besides, MADRL-based beamforming for noncoherent joint transmission is studied, where multi-antenna BSs jointly serve single-antenna UEs \cite{bai2024distributed}.
However, these schemes can not be directly applied to the non-joint transmission scenario, where one UE can be served by only one BS.
To the best of our knowledge, research on DRL-based UA and beamforming scheme remains limited in the non-joint transmission scenario.

Safe DRL has been studied for various applications due to its ability to deal with decision problems with safety constraints \cite{li2019constrained, wu2024real, zhao2024safe, yu2024causal}.
In \cite{li2019constrained}, the safe DRL-based electric vehicles charging scheduling is studied, where the cost measures the deviation of battery energy from the charging target.
In \cite{wu2024real}, safe DRL is applied to the real-time optimal power flow problem.
In \cite{zhao2024safe}, safe DRL is applied to the UAV-aided task offloading with the energy consumption constraint of the UAV.
In \cite{yu2024causal}, safe DRL is exploited to maximize the sum rate under the average rate per user constraint in the UAV-enabled wireless network.
Nonetheless, the application of safe DRL in wireless communication is still rare, and the aforementioned studies are limited to a single agent.
To this end, in this study, we design a safe DRL-based DDUACBF framework for the CATNs to maximize the sum rate of the TUs while ensuring the protection of the AUs.

\section{System Model}
\label{secSystemModel}
\begin{figure}[t]
\centering
\includegraphics[width=0.8\linewidth]{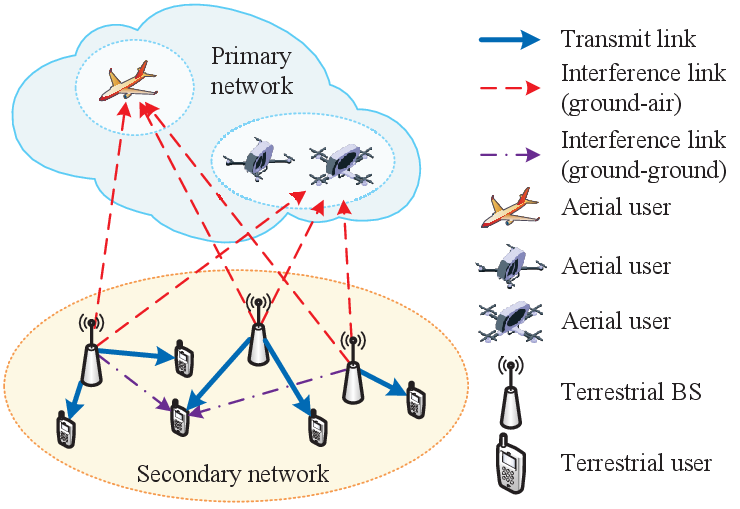}
\caption{Cognitive aerial-terrestrial network.}
\label{scenario}
\vspace{-0.4cm}
\end{figure}

As shown in Fig. \ref{scenario}, we consider a CATN, where the aerial network acts as a primary network to share its spectrum with the secondary terrestrial network if the interference from the terrestrial network can be tolerated by the AUs.
In the aerial network, there are $L$ AUs receiving downlink signals.
In the terrestrial network, there are $N$ BSs and $K$ single-antenna TUs.
Each TU can only receive downlink signals from one associated BS.
Each BS is equipped with a uniform rectangular array (URA) with $M = M_h \times  M_v$ antennas, where $M_h$ and $M_v$ are the number of horizontal and vertical antennas, respectively.
Besides, we denote the set of the indices of all AUs, the set of the indices of all BSs, and the set of the indices of all TUs by ${\mathcal U}^{\rm{pri}} \triangleq  \left\{ 1, \cdots ,L \right\}$, ${\mathcal B} \triangleq  \left\{ {1, \cdots ,N} \right\}$, and ${\mathcal U} \triangleq  \left\{ {1, \cdots ,K} \right\}$, respectively.
At the beginning of each time slot, each TU needs to decide which BS to associate with.
The BS associated with TU $k$ in time slot $t$ is denoted as $\varrho_k\left( t \right) \in {\cal B}$.
If handover occurs in time slot $t$, i.e., BS $\varrho_k\left( t \right)$ is different from BS $\varrho_k\left( t-1 \right)$, then TU $k$ should spend part of the time slot for handover overhead.
%
Accordingly, the set of indices of TUs associated with BS $n$ is denoted as ${{\cal K}_n}\left( t \right) = \left\{ {k|{\varrho _k}\left( t \right) = n} \right\}$.
In the following, we omit the time slot index for brevity if there is no misunderstanding.

\subsection{Transmission Model}

The received signal at TU $k$ can be expressed as:
\begin{align}
{y_k} = {\bf{h}}_{\varrho_k,k}^H{{\bf{w}}_k}{x_k} + \sum\limits_{i \in {\cal K},i \ne k} {{\bf{h}}_{{\varrho _i},k}^H{{\bf{w}}_i}{x_i}}  + {z_k}, \nonumber
\end{align}
where $x_{k}$ is the information signal for TU $k$, ${{\bf{w}}_k} \in \mathbb{C}^{M \times 1}$ is the beamforming vector of BS ${\varrho _k}$ for TU $k$, ${\bf{h}}_{n,k} \in \mathbb{C}^{M \times 1}$ is the channel from BS $n$ to TU $k$, and $z_{k}$ is the additive white complex Gaussian noise that follows ${\mathcal C}{\mathcal N} (0, \sigma _{k}^2)$.

Then, the signal-to-interference-plus-noise ratio (SINR) of TU $k$ for decoding $x_{k}$ can be expressed as: 
\vspace{-0.08cm}
\begin{align}
\label{SINR_kth_user}
{\gamma _k} = \frac{{{{\left| {{\bf{h}}_{\varrho_k,k}^H{{\bf{w}}_k}} \right|}^2}}}{{\sum\limits_{i \in {\cal K},i \ne k} {{{\left| {{\bf{h}}_{{\varrho _i},k}^H{{\bf{w}}_i}} \right|}^2}}  + \sigma _k^2}}.
\end{align}
Therefore, the achievable rate of TU $k$ can be expressed as ${R_k} = {\log _2}\left( {1 + {\gamma _k}} \right)$.

\subsection{Channel Model}
\label{secChannelModel}
%
The channels from BS $n$ to TU $k$ in time slot $t$ is modeled by the Rayleigh channel model, i.e., 
\begin{align}
    {{\bf{h}}_{n,k}}\left( t \right) = \sqrt {L_{{\rm{UMa}},n,k}^{ - 1}\left( t \right)} {\bf{h}}_{n,k}^{{\rm{NLoS}}}\left( t \right),\nonumber
\end{align}
where $L_{{\rm{UMa}},n,k}\left( t \right)$ is the path loss generated according to the Urban Macro (UMa) scenario of 3GPP TR 38.901 \cite{3GPP38901}. 
It is probabilistic LoS and is a function of the distance between BS $n$ and TU $k$ ${d_{n,k}}$ and the carrier frequency $f_c$.
Besides, the non-line-of-sight (NLoS) component is given by \cite{kim2011impact}:
\begin{align}
{\bf{h}}_{n,k}^{{\rm{NLoS}}}\left( {t + 1} \right) = \alpha {\bf{h}}_{n,k}^{{\rm{NLoS}}}\left( t \right) + \sqrt {1 - {\alpha ^2}} {{\bf{e}}_{n,k}}\left( t \right),\nonumber
\end{align}
where $\alpha$ is the correlation coefficient of the Rayleigh fading vector between adjacent time slots, ${\bf{h}}_{n,k}^{{\rm{NLoS}}}\left( 0 \right)\sim {\mathcal C}{\mathcal N}\left( {0,{{\bf{I}}_M}} \right)$, and ${{\bf{e}}_{n,k}}\left( t \right)\sim {\mathcal C}{\mathcal N}\left( {0,{{\bf{I}}_M}} \right)$.

The channel from BS $\!n$ to AU $\!l$ in time slot $\!t$ is modeled by the Rician channel model, i.e., 
\begin{align}
{{\bf{g}}_{n,l}}\left( t \right) \!= \!\!\sqrt {L_{{\rm{FSP}},n,l}^{ - 1} \!\left( t \right)} \!\left(\! {\sqrt {\!\frac{\kappa}{{\kappa \!+\! 1}}} {\bf{g}}_{n,l}^{{\rm{LoS}}}\!\left( t \right) + \!\sqrt {\!\frac{1}{{\kappa \!+\! 1}}} {\bf{g}}_{n,l}^{{\rm{NLoS}}}\!\left( t \right)} \!\right)\!,\nonumber
\end{align}
where $\kappa$ is the Rician factor and $L_{{\rm{FSP}},n,l}(t)$ is the path loss generated under the free space propagation model according to 3GPP TR 38.876 \cite{3GPP38876}.
$L_{{\rm{FSP}},n,l}(t)$ is a function of the distance between BS $n$ and AU $l$ $d_{n,l}(t)$ and $f_c$. 
%
%
Moreover, the LoS component is given by:
\begin{align}
{\bf{g}}_{n,l}^{{\rm{LoS}}}\left( t \right) = {e^{ - j2\pi d_{n,l}(t)/\lambda }}{\bf{a}}\left( {{\theta _{n,l}}\left( t \right),{\phi _{n,l}}\left( t \right)} \right),\nonumber
\end{align}
where $2\pi d_{n,l}(t) f_c/c$ is the phase of the LoS path for the reference antenna and $c$ is the speed of light.
%
Besides, ${\theta _{n,l}}\left( t \right)$ and ${\phi _{n,l}}\left( t \right)$ are the elevation and azimuth angles of AU $l$ relative to BS $n$, respectively, and ${\bf{a}}\left( {\theta ,\phi } \right)$ is the steering vector, which is given by:
\begin{align}
{\bf{a}}\left( {\theta ,\phi } \right) = \frac{1}{{\sqrt M }}[&0, \ldots ,{e^{j\frac{{2\pi }}{\lambda }\Delta d\left( {\left( {{m_h} - 1} \right)\sin \theta \sin \phi  + \left( {{m_v} - 1} \right)\cos \theta } \right)}}, \nonumber\\
&\ldots ,{e^{j\frac{{2\pi }}{\lambda }\Delta d\left( {\left( {{M_h} - 1} \right)\sin \theta \sin \phi  + \left( {{M_v} - 1} \right)\cos \theta } \right)}}]^T,\nonumber
\end{align}
where $\Delta d$ is the horizontal/vertical antenna spacing.
%
Moreover, the NLoS component is given by: 
\begin{align}
{\bf{g}}_{n,l}^{{\rm{NLoS}}}\left( {t + 1} \right) = \alpha {\bf{g}}_{n,l}^{{\rm{NLoS}}}\left( t \right) + \sqrt {1 - {\alpha ^2}} {{\bf{e}}_{n,l}}\left( t \right),\nonumber
\end{align}
where ${\bf{g}}_{n,l}^{{\rm{NLoS}}}\left( 0 \right)\sim {\mathcal C}{\mathcal N}\left( {0,{{\bf{I}}_M}} \right)$ and ${{\bf{e}}_{n,l}}\left( t \right)\sim {\mathcal C}{\mathcal N}\left( {0,{{\bf{I}}_M}} \right)$.


\section{Problem Formulation and Optimization-based Scheme}
\label{secProblem}
\subsection{Problem Formulation}
As mentioned in Section \ref{secChannelModel}, we consider that the channel coefficients vary across different time slots while maintaining temporal correlation.
To obtain a high sum rate under changing channel conditions, the TUs need to select an appropriate BS to associate with and the BSs need to dynamically adjust their beamforming strategies.
Specifically, by optimizing the user association between TUs and BSs and the transmit beamforming vectors of the BSs, we aim to maximize the sum rate of the terrestrial network subject to the interference temperature constraints of the AUs and the maximum transmit power constraints of the BSs.
The optimization problem can be formulated as:  
\begin{subequations}
\label{Problem_ori}
\begin{align}
\label{Problem_ori_obj}
&{\max _{\left\{ {{\varrho _k}\left( t \right),{{\bf{w}}_k}\left( t \right)} \right\}}}\;
\sum\limits_{k = 1}^K {{{\log }_2}\left( {1 + {\gamma _k}\left( t \right)} \right)}\\
\label{cons_power}
&\;\;\;\;\;\;{\rm{s.t.}}\;\;\;\;\;\;
\sum\limits_{k \in {{\cal K}_n}} {\left\| {{{\bf{w}}_k}\left( t \right)} \right\|_2^2}  \le {P_{\max }},\forall n \in {\cal B},\\
\label{cons_interfer}
&\;\;\;\;\;\;\;\;\;\;\;\;\;\;\;\;\;\;
\sum\limits_{k = 1}^K {{{\left| {{\bf{g}}_{{\varrho _k}\left( t \right),l}^H\left( t \right){{\bf{w}}_k}\left( t \right)} \right|}^2}} \! \le {I_{\max }},\forall l \!\in {{\cal U}^{{\rm{pri}}}},\\
&\;\;\;\;\;\;\;\;\;\;\;\;\;\;\;\;\;\;
{\varrho _k}\left( t \right) \in {\cal B},
\end{align}
\end{subequations}
where ${P_{\max}}$ is the maximum transmit power of the BSs and ${I_{\max}}$ is the received interference power threshold of the AUs, i.e., the interference temperature limit.
%
Note that the beamforming vector can be determined by two parts, i.e., ${{\bf{w}}_k} = \sqrt {p_k} {\overline {\bf{w}} _k}$, where ${\overline {\bf{w}} _k} = {{\bf{w}}_k}/{\left\| {{{\bf{w}}_k}} \right\|_2}$ is the normalized part of ${{\bf{w}}_k}$ and ${p_k} = \left\| {{{\bf{w}}_k}} \right\|_2^2$ is the power part of ${{\bf{w}}_k}$.

\subsection{Optimization-based DCD-WMMSE Scheme}
\label{OptimizationAlgorithm}
Problem \eqref{Problem_ori} is NP-hard, which makes it difficult to find its optimal solution.
To find the near-optimal solution, we develop a two-stage UA and CBF scheme based on iterative optimization, which is a slightly modified version of the algorithm proposed in \cite{shen2014distributed}.
Specifically, in the first stage, the UA and the power control are iteratively optimized via the dual coordinate descent (DCD) algorithm in Algorithm \ref{alg_DCD} and the Newton's method, respectively \cite{shen2014distributed}.
The main idea of the DCD algorithm is that each user associates with a BS to maximize its utility minus the price, while the BSs choose their prices iteratively to balance their loads.
Note that the DCD algorithm requires the transmit power of all BSs $p_n, \forall n$ and the real-time global channel strength $\left\|{\bf{h}}_{n,k}\right\|^2, \forall n,k$.
Therefore, to obtain the user association variables $\varrho_k, \forall k$, this joint UA and power control algorithm incurs high information exchange overhead and computational complexity.

\begin{algorithm}[h] 
\small
\caption{Dual coordinate descent (DCD) algorithm}
\label{alg_DCD}
\begin{algorithmic}[1]
\STATE \textbf{Input:} the transmit power of all BSs $q_n, \forall n$ and all the channel strength $\left\|{\bf{h}}_{n,k}\right\|^2, \forall n,k$.
\STATE Calculate the utility of the TU $k$ if it is associated with BS $n$ by ${\overline{u}}_{n,k}=\log(M\log_2(1+{\mathrm{SINR}}_{n,k}))$, $\forall n,k$, where ${\mathrm{SINR}}_{n,k}=\frac{|{\bf{h}}_{n,k}|^2 q_n}{\sum_{m\neq n}{\left|{\bf{h}}_{m,k}\right|^2 q_m}+\sigma_k^2}$.
\STATE Set the dual variable, i.e., the price at BS $n$ ${\overline{\mu}}_n=0, \forall n$ and the dual variable ${\overline{\nu}}  = \log \left( {\frac{1}{K}\sum_n {{e^{{{\overline{\mu}} _n} - 1}}} } \right)$.
\REPEAT
\FOR{$n \in \left\{ {1, \cdots ,N} \right\}$} 
\STATE Update ${\overline{\mu}} _n^{(t + 1)} = \sup \!\left\{ {{{\overline{\mu}} _n}|f_2^{(t)}\!\left( {{{\overline{\mu}} _n}} \right) - f_1^{(t)}\!\left( {{{\overline{\mu}} _n}} \right) \!\le\! 0} \right\}$, where ${f_2}\left( {{{\overline{\mu}} _n}} \right) = {e^{{{\overline{\mu}} _n} - {\overline{\nu}}  - 1}}$, ${f_1}\left( {{{\overline{\mu}} _n}} \right) = \left| {{{\mathcal U}_n}} \right|$, and ${{\mathcal U}_n} = \left\{ {k|{{\overline{u}}_{n,k}} - {{\overline{\mu}} _n} = \mathop {\max }_m \left( {{{\overline{u}}_{m,k}} - {{\overline{\mu}} _m}} \right)} \right\}$.
\ENDFOR
\STATE Update ${\overline{\nu}} {^{(t + 1)}} = \log \left( {\frac{1}{K}\sum_n {{e^{{\overline{\mu}} _n^{(t)} - 1}}} } \right)$.
\UNTIL the dual objective value $g\left( {{\overline{\boldsymbol{\mu}}} ,{\overline{\nu}} } \right)$ converges, where $g\left( {{\overline{\boldsymbol{\mu}}} ,{\overline{\nu}} } \right) = \sum_k {\mathop {\max }_n } \left( {{{\overline{u}}_{n,k}} - {{\overline{\mu}} _n}} \right) + \sum_n {{e^{{{\overline{\mu}} _n} - {\overline{\nu}}  - 1}}}  + {\overline{\nu}} K$.
\STATE Set ${\varrho _k} = {{\mathop{\rm argmax}\nolimits} _m}\left( {{{\bar u}_{m,k}} - {{\bar \mu }_m}} \right),\forall k$.
\RETURN the user association variables $\varrho_k, \forall k$.
\end{algorithmic}
\end{algorithm}

In the second stage, the beamforming vectors are designed based on the idea of the WMMSE algorithm \cite{shi2011iteratively}, as detailed in Algorithm \ref{alg_Closed_Form_FP_cog}.
However, this iterative optimization-based algorithm requires real-time global CSI and exhibits high computational complexity, rendering it unsuitable for deployment in CATN, which includes massive cellular networks.

\begin{algorithm}[h] 
\small
\caption{WMMSE-based CBF algorithm}
\label{alg_Closed_Form_FP_cog}
\begin{algorithmic}[1]
\STATE Initialize ${{\bf{w}}_k}, \forall k$, such that $\sum_{k \in {{\mathcal K}_n}} \!{\left\| {{{\bf{w}}_k}} \right\|_2^2}  \!=\! {P_{\max }}, \forall n$.
\REPEAT 
\STATE ${u_k} = \frac{{{{\left| {{\bf{h}}_{{\varrho _k},k}^H{{\bf{w}}_k}} \right|}^2}}}{{\sum\limits_{i \in {\cal K},i \ne k} {{{\left| {{\bf{h}}_{{\varrho _i},k}^H{{\bf{w}}_i}} \right|}^2}}  + \sigma _k^2}}$, $\forall k$.
\STATE ${v_k} = \frac{{\sqrt {(1 + {u_k})} {\bf{h}}_{{\varrho _k},k}^H{{\bf{w}}_k}}}{{\sum\limits_{i \in {\cal K}} {{{\left| {{\bf{h}}_{{\varrho _i},k}^H{{\bf{w}}_i}} \right|}^2}}  + \sigma _k^2}}$, $\forall k$ and ${\alpha _k} = {\left| {{v_k}} \right|^2}$.
\STATE ${{\bf{w}}_k} = {\bf{D}}_{{\varrho _k}}^{ - 1}\sqrt {(1 + {u_k})} {{\bf{h}}_{{\varrho _k},k}}{v_k}$, $\forall k$, where ${{\bf{D}}_n} = \sum_{i = 1}^K {{\alpha _i}{{\bf{h}}_{n,i}}{\bf{h}}_{n,i}^H}   + \sum_{l = 1}^L {{\mu _l}{{\bf{g}}_{n,l}}{\bf{g}}_{n,l}^H}  + {\eta _n}{\bf{I}}$.
\UNTIL the objective function \eqref{Problem_ori_obj} converges.
\RETURN ${{\bf{w}}_k}, \forall k$.
\end{algorithmic}
\end{algorithm}

\section{Safe DRL-based Distributed Dynamic User Association and Coordinated Beamforming in CATN}
\label{secproposedScheme}
According to the above analysis, the real-time global CSI is still required for the traditional optimization-based algorithms to obtain the optimal UA and beamforming.
To alleviate the strict requirements of CSI, we develop a DDUACBF scheme, where each TU or BS makes decisions based on local information and information obtained from other TUs or BSs, rather than relying on complete knowledge of the wireless environment.
Considering the delay of information transmission, part of the obtained non-local information is only available in the next time slot.
To optimize the agent's decision-making in an unknown environment, DRL algorithms can be exploited.


In this section, we first introduce the NCPOMG and safe reinforcement learning (RL).
Then, we show that in CATN, the considered problem can be described as an NCPOMG.
Subsequently, we present the workflow of the proposed safe DRL-based DDUACBF framework, where each TU is an agent for UA and each BS is an agent for beamforming.
We then elaborate on the composition of the BS agents and the TU agents, including the algorithms they use.
Finally, we summarize the safe DRL-based DDUACBF scheme.

\subsection{Preliminaries of NCPOMG and Safe Multi-Agent RL}
A NCPOMG model for $N$ agents can be described with a tuple $\left\langle {{\mathcal N},{\mathcal S},\Omega, {\mathcal A},{\mathcal P},{\rho ^0},{\mathcal R},{\mathcal C},{\mathcal B},{\gamma _r},{\gamma _c}, {\left\{ {{{\cal G}_t}} \right\}}_{t \ge 0}} \right\rangle$, whose elements are respectively defined as follows \cite{zhang2021decentralized, feriani2021single, gu2023safe}:
\begin{itemize}
\item ${\mathcal N} = \left\{ {1, \ldots ,N} \right\}$ is the set of agents.
\item ${\mathcal S}$ denotes the state space.
\item $\Omega = \prod_{n = 1}^N {\Omega_n}$ denotes the joint observation space, where ${\Omega_n}$ is the set of observations for agent $n$ and ${\bf{o}} = \left\langle {{{\bf{o}}_1}, \cdots ,{{\bf{o}}_N}} \right\rangle  \in \Omega$ denotes the joint observation.
\item ${\mathcal A} = \prod_{n = 1}^N {{\mathcal{A}}_n}$ denotes the joint action space, where $\mathcal{A}_n$ denotes the set of actions for agent $n$ and ${\bf{a}} = \left\langle {{{\bf{a}}_1}, \cdots ,{{\bf{a}}_N}} \right\rangle  \in {\mathcal A}$ denotes the joint action.
\item $P\left( {{\bf{s}}^{\prime}|{\bf{s}},{\bf{a}}} \right) \in {\mathcal P}$ is the probabilistic  transition function.
\item ${\rho ^0}$ is the initial state distribution.
\item ${\mathcal{R}}=\left\{r_n \right\}_{n\in {\mathcal{N}}}$ is the set of reward functions, where $r_n :{\mathcal S} \times {\mathcal A}_n \times {\mathcal S} \to \mathbb{R}$ is the reward function for agent $n$.
\item ${\mathcal{C}}=\left\{c_{n,l}\right\}_{n\in {\mathcal{N}}, 1\leq l\leq L_n}$ is set of cost functions, where $c_{n,l}:{\mathcal S} \times {\mathcal A}_n \times {\mathcal S} \to \mathbb{R}$ is the $l$-th cost function for agent $n$ and agent $n$ has $L_n$ cost functions.
\item ${\mathcal{B}}=\left\{b_{n,l}\right\}_{n\in {\mathcal{N}}, 1\leq l\leq L_n}$ is the set of corresponding cost limits, where $b_{n,l}$ is the $l$-th cost limit for agent $n$.
\item $\gamma_r,\gamma_c\in(0,1)$ are the discount factor for reward and cost.
\item ${{\cal G}_t}{\rm{ = }}\left( {{\cal N},{{\cal E}_t}} \right)$ is the time-varying communication network at time $t$, which links $N$ nodes with edges in ${{\cal E}_t}$. An edge $(i, j) \in {{\cal E}_t}$, $\forall i, j$ means that agents $i$ and $j$ can communicate mutually at time $t$.
\end{itemize}

Denote the policy for agent $n$ as $\pi_n:{\Omega_n}\times{\mathcal{A}}_n\to[0,1]$, ${\boldsymbol{\pi}}=\left\{\pi_n\right\}_{n\in {\mathcal{N}}}$, the expected discounted cumulative reward for agent $n$ is ${J_R}\left( {{\pi _n}} \right) = {{\mathbb{E}}_{{\bf{s}} \left( 0 \right),{{\bf{a}}_n}\left( 0 \right),...}}\left[ {\sum_{t = 0}^\infty  {\gamma _r^t{r_n}\left( t \right)} } \right]$.
NCPOMG aims to find an optimal joint policy $\boldsymbol{\pi}^*$ that maximizes the average ${J_R}\left( {{\pi _n}} \right)$ of all agents ${J_R}\left( {\boldsymbol{\pi}} \right) \buildrel \Delta \over = \frac{1}{N}\sum_{n \in {\mathcal N}} {{J_R}\left( {{\pi _n}} \right)}$ while satisfying the safety constraints, NCPOMG can be formulated as:
\begin{align}
\label{NCPOMG}
{{\boldsymbol{\pi}} ^{*}}=\mathop {{\rm{argmax}}}_{\left\{ \pi_n  \in {\Pi _{{\mathcal C}, n}} \right\}_{n\in {\mathcal{N}}} } {J_R}\left( {\boldsymbol{\pi}} \right),
\end{align}
where ${\Pi _{{\mathcal C}, n}} \triangleq \left\{ \pi_n  \in \Pi_n :{J_{C,n,l}}\left( \pi_n  \right) \le {b_{n,l}}, \forall l = 1,...,L_n \right\}$ is the feasible policy set, ${J_{C,n,l}}\left( {{\pi _n}} \right) \buildrel \Delta \over = {{\mathbb{E}}_{{{\bf{s}}\left( 0 \right)},{{\bf{a}}_n\left( 0 \right)},...}}\left[ {\sum_{t = 0}^\infty  {\gamma _c^t{c_{n,l}}\left( t \right)} } \right]$ is the $l$-th expected discounted cumulative cost for agent $n$.
The inequalities ${J_{C,n,l}}\left( \pi_n  \right) \le {b_{n,l}}, \forall l$ in ${\Pi _{{\mathcal C}, n}}$ are the safety constraints for agent $n$.
Besides, ${\bf{s}}\left( 0 \right) \sim{\rho ^0}$, ${\bf{a}}_n\left( t \right) \sim{\pi _n}\left( { \cdot |{\bf{o}}_n\left( t \right)} \right)$, and ${\bf{s}}\left( t+1 \right)\sim P\left( { \cdot |{\bf{s}}\left( t \right),{{\bf{a}}\left( t \right)}} \right)$.
To find the optimal joint policy $\boldsymbol{\pi}^*$, safe multi-agent RL algorithms has been studied \cite{gu2023safe}.


In multi-agent systems, to achieve a joint goal, effective coordination among agents is essential to make the joint action optimize the performance of mutual tasks.
However, partial observability and non-stationarity pose challenges. 
Specifically, a single agent can only observe partial information about the state of the environment. 
In addition, when multiple agents interact and learn simultaneously in a shared environment, the environment may be non-stationary from the perspective of any individual agent.
To address these challenges, facilitating information exchange among agents proves to be an effective solution.
Specifically, the networked agents in decentralized multi-agent RL are allowed to exchange information with their neighbors over a communication network \cite{zhang2021decentralized, feriani2021single}. 
Based on the collected information and the local observations, each agent makes its own decision without the coordination of a central controller.


\subsection{NCPOMG-based Framework and its Workflow}
\begin{figure*}[t]
\centering
\includegraphics[width=0.8\linewidth]{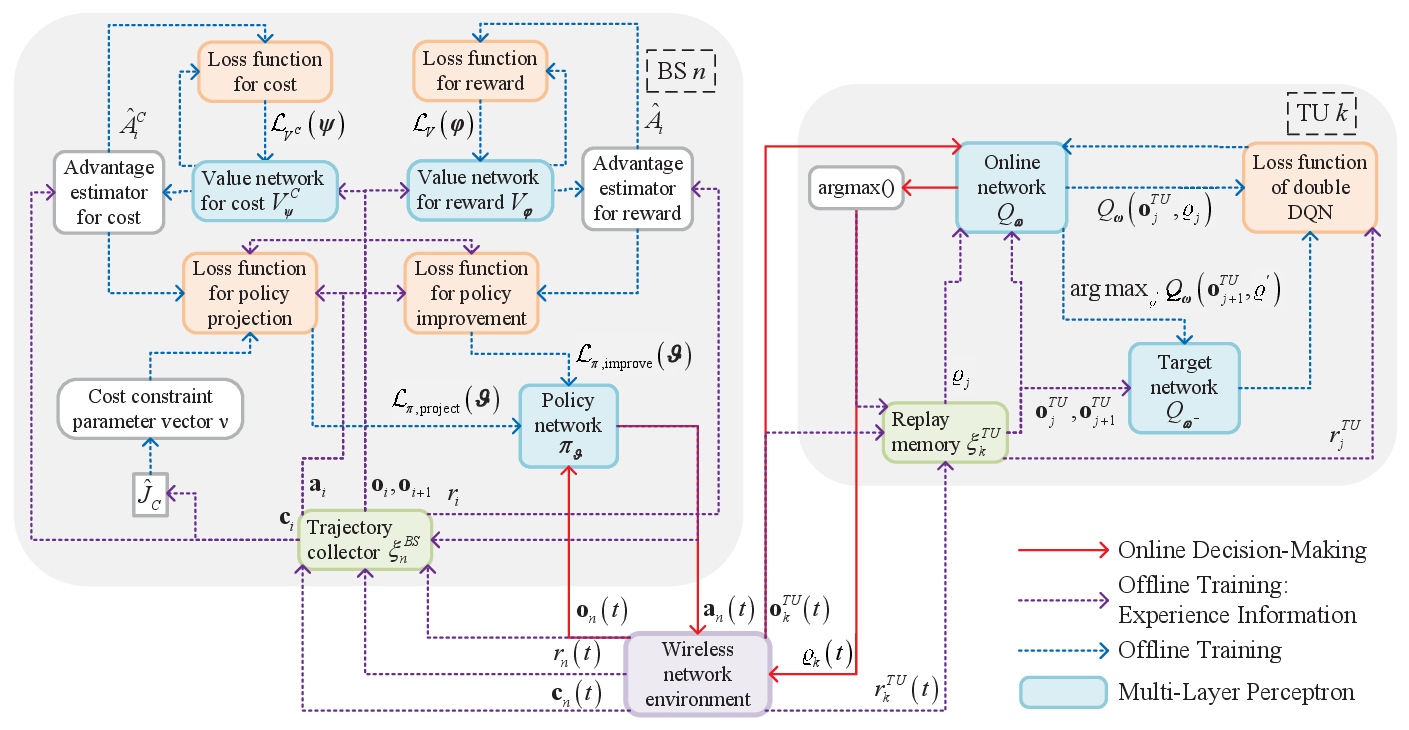}
\caption{Illustration for the workflow of the proposed safe DRL-based DDUACBF framework.}
\label{workflow_DRL}
\vspace{-0.4cm}
\end{figure*}

In CATN, the BSs and the TUs can act as agents that observe the wireless network environment and then make decisions based on the observed information to achieve goals such as maximizing the rate.
As indicated in \eqref{SINR_kth_user}, the optimal beamforming decision that maximizes the sum rate of TUs associated with a BS is affected by the beamforming decision of other BSs.
Thus, the information observed by an agent at time $t+1$ should contain information about the decisions made by other agents at time $t$.
This process can be modeled as a Markov game.
Since each agent cannot observe complete information about the environment, the system can be modeled as a partially observable Markov game (POMG) \cite{feriani2021single}.
In particular, to maximize the sum rate in \eqref{SINR_kth_user}, it is not enough for the BS to rely solely on the local CSI (i.e., the CSI from it to all TUs and AUs) and it is necessary to acquire CSI from other BSs.
To mitigate the challenge of partial observability, the agents are considered to be networked, i.e., they can exchange information through a communication network \cite{zhang2021decentralized}.
Moreover, for BS agents, their beamforming decisions need to satisfy the safe constraints related to the received interference power of the AUs. 
To this end, after making a decision, each BS agent will receive a cost vector that indicates how safe the decision is \cite{gu2023safe}.
In summary, the considered UA and beamforming problem in CATN can be modeled as NCPOMG, which can be solved by advanced safe DRL techniques.

Fig. \ref{workflow_DRL} shows the workflow of the proposed safe DRL-based DDUACBF framework, where all agents work in the decentralized training and decentralized execution (DTDE) way with information exchange by communication.
Specifically, each TU is a UA agent to decide which BS to associate with and each BS is a beamforming agent to decide the beamforming vectors for its associated TUs.
The composition of each agent is elaborated in the subsequent two subsections.


\subsection{BS Agent for Beamforming}
The adopted safe DRL algorithm, observation space, action space, reward function, and cost function of the BS agents are detailed as follows.

\subsubsection{Algorithm}
We apply a safe DRL algorithm, constrained update projection (CUP) \cite{yang2022constrained}, in each BS agent.
The CUP algorithm, which is on-policy and model-free, is one of the policy optimization-based approaches \cite{gu2024review}.
Compared with the constrained policy optimization (CPO) algorithm \cite{achiam2017constrained}, which is the first model-free policy gradient method to solve the CMDP problem and uses second-order proximal optimization, the CUP algorithm utilizes only first-order optimization, resulting in lower computational complexity.
Specifically, a CUP agent comprises three main parts: the actor, the reward critic, and the cost critic.
These three parts are actually the policy network $\pi_{\boldsymbol{\vartheta}}$, the value network for reward $V_{\boldsymbol{\varphi}}$, and the value network for costs $V^C_{\boldsymbol{\psi}}$ with parameters ${\boldsymbol{\vartheta}}$, ${\boldsymbol{\varphi}}$, and ${\boldsymbol{\psi}}$, respectively.
The policy for each agent is represented by ${\pi _{\boldsymbol{\vartheta}} }\left( {\bf{a}}|{\bf{o}} \right)$, which denotes the probability of choosing ${\bf{a}}$ in observation ${\bf{o}}$. 
Moreover, the observation ${\bf{o}}$ is evaluated by the value network for reward with the V value ${V_{\boldsymbol{\varphi}}}({\bf{o}})$ and the value network for costs with the V values ${V_{\boldsymbol{\psi }}}({\bf{o}})$.

Based on the collected $B_m^{BS}$ experiences, the BS agent updates its neural networks with the CUP algorithm, i.e., Algorithm \ref{alg_CUP}.
Every time the agent updates the networks, $B_b$ experiences are sampled to form a minibatch.
The parameters ${\boldsymbol{\varphi}}$ and ${\boldsymbol{\psi}}$ are updated by minimizing the mean square error (MSE) of the value network output and the target value.
Specifically, the loss functions of $V_{\boldsymbol{\varphi}}$ and $V^C_{\boldsymbol{\psi}}$ are given by:
\begin{align}
\label{value_network_reward_loss_function}
{{\mathcal L}_V}({\boldsymbol{\varphi}} ) &= \frac{1}{B_b}\sum_{j = 1}^{B_b} {{{\left( {{V_{\boldsymbol{\varphi}} }\left( {{{\bf{o}}_j}} \right) - V_j^{{\rm{target}}}} \right)}^2}},\\
\label{value_network_cost_loss_function}
{{\mathcal L}_{{V^C}}}({\boldsymbol{\psi}} ) &= \frac{1}{B_b}\sum_{j = 1}^{B_b} {{{\left( {{V^C_{\boldsymbol{\psi}} }\left( {{{\bf{o}}_j}} \right) - V_j^{C,{\rm{ target }}}} \right)}^2}}.
\end{align}
%
%
The policy is updated in a two-step approach, which contains performance improvement and projection \cite{yang2022constrained}.
Firstly, CUP performs a policy improvement, which may produce a temporary policy that violates the constraint. 
Secondly, the CUP algorithm projects the policy back onto the safe region to reconcile the constraint violation.
The gradients to update the parameters of the policy network are provided in \eqref{CUP_update_policy_Improvement} and \eqref{CUP_update_policy_Projection} at the top of this page, where the KL divergence between ${{\pi _{\boldsymbol{\vartheta}} }}$ and ${{\pi _{{{\boldsymbol{\vartheta}} ^\prime }}}}$ is denoted by ${D_{{\rm{KL}}}}\left( {{\pi _{\boldsymbol{\vartheta}} },{\pi _{{{\boldsymbol{\vartheta}} ^\prime }}}} \right)\left[ {\bf{o}} \right]$, i.e., ${D_{{\rm{KL}}}}\left( {{\pi _{\boldsymbol{\vartheta}} }\left( { \cdot |{\bf{o}}} \right),{\pi _{{{\boldsymbol{\vartheta}} ^\prime }}}\left( { \cdot |{\bf{o}}} \right)} \right)$.
Besides, $\hat{A}_j$ and $\hat{A}_j^C$ are the advantage functions estimated by the generalized advantage estimator (GAE) with parameter $\lambda$ \cite{yang2022constrained},
${\boldsymbol{\nu}} = {\left[ {{\nu _1}, \cdots ,{\nu _L}} \right]^T}$ is a cost constraint parameter vector.
%
We give a rough understanding about the policy update: the policy improvement in \eqref{CUP_update_policy_Improvement} aims to maximize the advantage function for reward without updating too much, and then the policy projection in \eqref{CUP_update_policy_Projection} aims to minimize the weighted sum of the policy difference and the advantage function for cost with the weight ${\boldsymbol{\nu}}$.

\begin{algorithm}[t] 
\small
\caption{Constrained update projection (CUP) algorithm}
\label{alg_CUP}
\begin{algorithmic}[1]
\STATE \textbf{Input:} experiences $\langle {{{\bf{o}}_{i}},{{\bf{a}}_{i}},{r_{i}},{{\bf{o}}_{i + 1}},{{\bf{c}}_{i}}} \rangle$, $i = 1, \ldots, B_m^{BS}$.
\STATE Estimate advantage functions ${{\hat A}_{i}}$ by ${{\hat A}_{i}} = \gamma \lambda {{\hat A}_{i + 1}} + {\delta _{i}}$, where $\hat A_{ B_m + 1} = 0$ and ${\delta _{i}} = {r_{i}} + \gamma {V_{\boldsymbol{\varphi}} }\left( {{{\bf{o}}_{i + 1}}} \right) - {V_{\boldsymbol{\varphi}} }\left( {{{\bf{o}}_{i}}} \right)$.
\STATE Estimate advantage functions $\hat A_{i}^C$ by $\hat A_{i}^C = \gamma \lambda \hat A_{i + 1}^C + \delta _{i}^C$, where $\hat A_{ B_m + 1}^C = 0$ and $\delta _{i}^C = {{\bf{c}}_{i}} + \gamma V^C_{\boldsymbol{\psi}} \left( {{{\bf{o}}_{i + 1}}} \right) - V^C_{\boldsymbol{\psi}} \left( {{{\bf{o}}_{i}}} \right)$.

\STATE Get $V_{i}^{\rm{ target }} = {{\hat A}_{i}} + V_{\boldsymbol{\varphi}} \left( {{{\bf{o}}_{i}}} \right)$ and $V_{i}^{C,{\rm{ target }}} = \hat A_{i}^C + V^C_{\boldsymbol{\psi}} \left( {{{\bf{o}}_{i}}} \right)$.
\STATE Estimate $C$-return ${{\hat J}_C}$ by ${{\hat J}_C} = \frac{1}{B_m^{BS}}\sum_{i = 1}^{B_m^{BS}}  {{\bf{c}}_{i}}$.

\STATE Update ${\boldsymbol{\nu}}$ by ${\boldsymbol{\nu}}  \leftarrow {\rm{clip}}\left( {{\boldsymbol{\nu}}  + {\alpha _{\boldsymbol{\nu}} }\left( {{\hat J}_C} - {\bf{b}} \right), 0, {{\boldsymbol{\nu}} _{\max }}} \right)$.

\STATE \textbf{Step 1: Performance Improvement}
\STATE Store old policy ${\boldsymbol{\vartheta}}^{\prime} \leftarrow {\boldsymbol{\vartheta}}$.
\FOR{$B_e$ epochs}
\FOR{each minibatch $\langle {{{\bf{o}}_j},{{\bf{a}}_j},{{\hat A}_j},{\hat A}_j^C,V_j^{{\rm{target }}},V_j^{C,{\rm{ target }}}} \rangle$ of size $B_b$}
\STATE Update ${\boldsymbol{\varphi}}  \leftarrow {\boldsymbol{\varphi}}  - {\alpha _V}{\nabla _{\boldsymbol{\varphi}} }{{\mathcal L}_V}({\boldsymbol{\varphi}} )$ with \eqref{value_network_reward_loss_function}.
\STATE Update ${\boldsymbol{\psi}}  \leftarrow {\boldsymbol{\psi}}  - {\alpha _V}{\nabla _{\boldsymbol{\psi}} }{{\mathcal L}_{{V^C}}}({\boldsymbol{\psi}} )$ with \eqref{value_network_cost_loss_function}.
\STATE Update ${\boldsymbol{\vartheta}}  \leftarrow {\boldsymbol{\vartheta}}  - {\alpha _\pi }{{\hat \nabla }_{\boldsymbol{\vartheta}} }{{\mathcal L}_{\pi,  {\rm{improve}}} }({\boldsymbol{\vartheta}} )$ with \eqref{CUP_update_policy_Improvement}.
\ENDFOR
\IF{$\frac{1}{B_m^{BS}}  {\sum_{i = 1}^{B_m^{BS}} {{D_{{\rm{KL}}}}\left( {{\pi _{\boldsymbol{\vartheta}} }, {\pi _{{{\boldsymbol{\vartheta}} ^\prime }}}} \right)\left[ {{{\bf{o}}_{i}}} \right]} }  > \varepsilon$} 
\STATE Break.
\ENDIF 
\ENDFOR

\STATE \textbf{Step 2: Projection}
\STATE Store old policy ${\boldsymbol{\vartheta}}^{\prime\prime} \leftarrow {\boldsymbol{\vartheta}}$.
\FOR{$B_e$ epochs}
\FOR{each minibatch $\langle {{{\bf{o}}_j},{{\bf{a}}_j},{{\hat A}_j},{\hat A}_j^C,V_j^{{\rm{target }}},V_j^{C,{\rm{ target }}}} \rangle$ of size $B_b$}
\STATE Update ${\boldsymbol{\vartheta}}  \leftarrow {\boldsymbol{\vartheta}}  - {\alpha _\pi }{{\hat \nabla }_{\boldsymbol{\vartheta}} }{{\mathcal L}_{\pi,  {\rm{project}}} }({\boldsymbol{\vartheta}} )$ with \eqref{CUP_update_policy_Projection}.
\ENDFOR
\IF{$\frac{1}{B_m^{BS}}  {\sum_{i = 1}^{B_m^{BS}} {{D_{{\rm{KL}}}}\left( {{\pi _{\boldsymbol{\vartheta}} }, {\pi _{{{\boldsymbol{\vartheta}} ^\prime }}}} \right)\left[ {{{\bf{o}}_{i}}} \right]} }  > \varepsilon$} 
\STATE Break.
\ENDIF 
\ENDFOR

\RETURN Policy network $\pi_{\boldsymbol{\vartheta}}$ and Value networks $V_{\boldsymbol{\varphi}}$, $V^C_{\boldsymbol{\psi}}$.
\end{algorithmic}
\end{algorithm}


\begin{figure*}[htbp]
\begin{align}
\label{CUP_update_policy_Improvement}
{{\mathcal L}_{\pi,  {\rm{improve}}}}({\boldsymbol{\vartheta}} ) =& -\frac{1}{B_b}\sum_{j = 1}^{B_b} {\min \left\{ {\frac{{{\pi _{\boldsymbol{\vartheta}} }\left( {{\bf{a}}_j\mid {\bf{o}}_j} \right)}}{{{\pi _{{{\boldsymbol{\vartheta}} ^{\prime}}}}\left( {{\bf{a}}_j\mid {\bf{o}}_j} \right)}}{{\hat A}_j},{\rm{clip}}\left( {\frac{{{\pi _{\boldsymbol{\vartheta}} }\left( {{\bf{a}}_j\mid {\bf{o}}_j} \right)}}{{{\pi _{{{\boldsymbol{\vartheta}} ^{\prime}}}}\left( {{\bf{a}}_j\mid {\bf{o}}_j} \right)}},1 - \varepsilon ,1 + \varepsilon } \right){{\hat A}_j}} \right\}},\\
\label{CUP_update_policy_Projection}
{{\mathcal L}_{\pi,  {\rm{project}}}}({\boldsymbol{\vartheta}} ) =& \frac{1}{B_b}\sum_{j = 1}^{B_b} \left( \nabla_{\boldsymbol{\vartheta}} D_{\rm{KL}}\left(\pi_{\boldsymbol{\vartheta}} , \pi_{{\boldsymbol{\vartheta}}^{\prime \prime}}\right)\left[{\bf{o}}_j\right] +  {\boldsymbol{\nu}}^T \frac{{1 - \gamma \lambda }}{{1 - \gamma }}\frac{{{\pi _{{\boldsymbol{\vartheta}}}}\left( {{\bf{a}}_j\mid {{\bf{o}}_j}} \right)}}{{{\pi _{{{{\boldsymbol{\vartheta}}}^{\prime}}}}\left( {{\bf{a}}_j\mid {{\bf{o}}_j}} \right)}}\hat A_{j}^C \right).
\end{align}
\hrule
\vspace{-0.4cm}
\end{figure*}

\subsubsection{Observation Space}
First, to reduce the number of elements of the state vector and the amount of information exchanged between BSs, we compress the channel ${\bf{h}}$ into a compressed channel ${{\bf{h}}^{\rm{c}}}$ by using a codebook ${\bf{F}} = [{{\bf{f}}_1}, \cdots ,{{\bf{f}}_C}] \in {\mathbb{C}}^{M \times C}$, where ${{\bf{f}}_c} \triangleq 1/\sqrt M \left[ {1,{e^{j2\pi c/C}}, \cdots ,{e^{j2\pi (M - 1)c/C}}} \right]$ and $C$ is the size of the codebook. 
Specifically, we calculate ${{\bf{F}}^H}{\bf{h}} = {\left[ {{d_1},{d_2}, \cdots ,{d_C}} \right]^T}$, whose elements are then sorted as $\left| {{d_{{c_1}}}} \right| \ge \left| {{d_{{c_2}}}} \right| \ge  \cdots  \ge \left| {{d_{{c_C}}}} \right|$.
Thus, given a compression factor $N_c$, we can obtain ${{\bf{h}}^{\rm{c}}} = \left[ {{c_1},{d_{{c_1}}},{c_2},{d_{{c_2}}}, \cdots ,{c_{{N_c}}},{d_{{c_{{N_c}}}}}} \right]$.

Then, we define some notations for the desired signal and interference information.
For TU $k$, the received desired signal power is $p_k^{\rm{r}} = {p_k}{\left| {{\bf{h}}_{{\varrho _k},k}^H {\overline {\bf{w}}_k}} \right|^2}$, the received interference power from BS $j$ is ${\beta _{j,k}} = \sum_{i \in {{\mathcal K}_j},i \ne k} {{p_i}{{\left| {{\bf{h}}_{j,k}^H{{\overline {\bf{w}}_i}}} \right|}^2}}$, and the received interference plus noise power is ${\beta _{k}} = \sum_{j = 1}^N {{\beta _{j,k}}}  + \sigma _{k}^2$.
Besides, if ${\varrho _k}\! \left( t \right) = n$, then the UA variable ${\chi _{n,k}}\! \left( t \right)=1$, otherwise ${\chi _{n,k}}\! \left( t \right)=0$.
In slot $t$, the local information of BS $n$ about the TUs in ${\mathcal{K}}_n\left( t \right)$ is:
\begin{align}
{\bf{o}}_n^{{\rm{loc}}}(t) = [ &{\boldsymbol{\chi}}_{n}^{\rm{BS}}(t), {\bf{H}}_n^{\rm{c}}(t),{{\bf{p}}_n}\left( {t - 1} \right),{{\bf{R}}_n}\left( {t - 1} \right),\nonumber\\
&{\bf{p}}_n^{\rm{r}}\left( {t - 1} \right),{{\boldsymbol{\beta}}_n}\left( {t - 1} \right) ],\nonumber
\end{align}
where ${\boldsymbol{\chi}}_{n}^{\rm{BS}}\triangleq[\chi_{n,1},\cdots,\chi_{n,K}]$, ${\bf{H}}_n^{\rm{c}} \triangleq {\boldsymbol{\chi}}_{n}^{\rm{BS}} \odot [ {\bf{h}}_{n,1}^{\rm{c}}, \cdots ,{\bf{h}}_{n,K}^{\rm{c}} ]$, ${{\bf{p}}_n} \triangleq {\boldsymbol{\chi}}_{n}^{\rm{BS}} \odot [ {p_1}, \cdots ,{p_K} ]$, ${\bf{R}}_n \triangleq {\boldsymbol{\chi}}_{n}^{\rm{BS}} \odot [ {R_1}, \cdots ,{R_K} ]$, ${\bf{p}}_n^{\rm{r}} \triangleq {\boldsymbol{\chi}}_{n}^{\rm{BS}} \odot [ p_1^{\rm{r}}, \cdots ,p_K^{\rm{r}} ]$, and ${{\boldsymbol{\beta }}_n} \triangleq {\boldsymbol{\chi}}_{n}^{\rm{BS}} \odot [ {\beta _{n,1}},\cdots ,{\beta _{n,K}} ]$.
Here the vector element-wise product is used to exclude non-local information.

The statistical knowledge of the channels includes the distance between the transmitter and the receiver, the pathloss parameter, and the array response vector.
Thus, the statistical information of channel from BS $n$ to AU $l$ is denoted by ${\bf{g}}_{n,l}^S = \left[ {{\theta _{n,l}},{\phi _{n,l}},L_{{\rm{FSP}},n,l}^{ - 1},{d_{n,l}}} \right]$.
For AU $l$, the inferred received interference power from BS $n$ is ${\rho _{n,l}} = \sum_{k\in{\mathcal{K}}_{n}}{p_k}\left|\sqrt{L_{{\rm{FSP}},n,l}^{-1}}\left({\bf{g}}_{n,l}^{LoS}\right)^{H}{\overline {\bf{w}}_k}\right|^{2}$ and the received interference power from all BSs is ${\rho _l} = \sum\limits_{k \in {\cal K}} {{p_k}{{\left| {{\bf{g}}_{{\varrho _k},l}^H{{\overline {\bf{w}} }_k}} \right|}^2}}$.
In slot $t$, the local information of BS $n$ about AUs is:
\begin{align}
{\bf{o}}_n^{{\rm{loc}},{\rm{pri}}}\left( t \right) = \left[ {\left[ {{\bf{g}}_{n,l}^S\left( t \right), {\rho _{n,l}}\left( {t - 1} \right)} \right]}_{l \in {{\mathcal U}^{{\rm{pri}}}}} \right].\nonumber
\end{align}

To describe the information obtained from other BSs, we first define the set of the interferer BSs of TU $k$ as ${\mathcal B}_{k}^{{\rm{in}}}(t) = \left\{ j \in {\mathcal B}\mid {\beta _{j,k}}(t) > {\varsigma _{k}}(t) \right\}$, where ${\varsigma _{k}}(t)$ is a threshold that ensures $\left| {{\mathcal B}_{k}^{{\rm{in}}}(t)} \right| = B_{\rm{in}}$.
%
Then, we define the information from ${\mathcal B}_{k}^{{\rm{in}}}(t - 1)$ as $\overline {\bf{o}}_{k}^{{\rm{in}}}(t) = \left[ {k, \left[ {j,{\bf{H}}_j^{\rm{c}}(t -\! 1),{{\bf{p}}_j}(t - \!1),{\beta _{j,k}}(t -\! 1)} \right]}_{j \in {\mathcal B}_{k}^{{\rm{in}}}(t - 1)} \right]$.
Thus, the information from the interferer BSs of some TUs in ${\mathcal{K}}_n\left( t-1 \right)$ is:
\begin{align}
{\bf{o}}_n^{{\rm{in}}}\left( t \right) = \left[ {{{\left[ {\overline {\bf{o}} _k^{{\rm{in}}}\left( t \right)} \right]}_{k \in {\mathcal U}_n^{{\rm{in}}}\left( {t - 1} \right)}}} \right],\nonumber
\end{align}
where ${\mathcal{U}}_n^\text{in}(t)=\{k\in{\mathcal{K}}_n\left( t \right) \mid \beta_k(t)>\bar{\varsigma}_n(t)\}$ is the set of the TUs in ${\mathcal{K}}_n\left( t \right)$ who suffer severe interference and $\bar{\varsigma}_n(t)$ is a threshold that ensures $\left|{\mathcal{U}}_{n}^{\rm{in}}(t)\right|=\min\{K_{\rm{in}}, \left| {{{\mathcal K}_n}\left( t \right)} \right|\}$.
If $\left| {{{\mathcal K}_n}\left( {t - 1} \right)} \right| < {K_{{\rm{in}}}}$, the empty information in ${\bf{o}}_n^{{\rm{in}}}\left( t \right)$ is complemented with zero so that the dimension of ${\bf{o}}_n^{{\rm{in}}}\left( t \right)$ is $\left((1+3N_{c}K+K+1)B_{\rm{in}}+1\right){K_{{\rm{in}}}}$.

Similarly, we define the set of the interferer BSs of AU $\!l$ as ${\mathcal B}_l^{{\rm{in,pri}}}(t) = \left\{ {j \in {\mathcal B}\mid {\rho _{j,l}}(t) > {\widetilde{\varsigma} _l}(t)} \right\}$, where ${\widetilde{\varsigma} _l}(t)$ is a threshold that ensures $\left| {{\mathcal B}_l^{{\rm{in,pri}}}(t)} \right| = \widetilde B_{\rm{in}}$.
%
Then, we define the information from ${\mathcal B}_l^{{\rm{in}},{\rm{pri}}}(t - 1)$ as ${\bf{o}}_l^{{\rm{in}},{\rm{pri}}}(t) = \left[ {{{\left[ {j,{\bf{g}}_{j,l}^S(t - 1),{{\bf{p}}_j}(t - 1),{\rho _{j,l}}(t - 1)} \right]}_{j \in {\mathcal B}_l^{{\rm{in}},{\rm{pri}}}(t - 1)}}} \right]$.
Thus, the information from the interferer BSs of all AUs is:
\begin{align}
{{\bf{o}}^{{\rm{in}},{\rm{pri}}}}(t) = \left[ {{{\left[ {{\bf{o}}_l^{{\rm{in}},{\rm{pri}}}\left( t \right)} \right]}_{l \in {{\mathcal U}^{{\rm{pri}}}}}}} \right].\nonumber
\end{align}

Then, we define the set of the interfered TUs of BS $n$ as ${\mathcal U}_n^{{\rm{out }}}(t) = \left\{ i \in {\mathcal U}\mid {\beta _{n,i}}(t) > {\varsigma _n}(t) \right\}$, where ${\varsigma _n}(t)$ is a threshold that ensures $\left| {{\mathcal U}_n^{{\rm{out }}}(t)} \right| = K_{\rm{out}} B_{\rm{in}}$.
If $\sum_{k = 1}^K {{\beta _{n,k}}} =0$, then ${\mathcal U}_n^{{\rm{out}}}\left( t \right) = \left\{ {i \in {\mathcal U}\mid \left\| {{{\bf{h}}_{n,i}}\left( t \right)} \right\|_2^2 > {\varsigma _n}\left( t \right)} \right\}$.
%
Then, we define the information from ${\mathcal U}_n^{{\rm{out}}}(t - 1)$ as:
\begin{align}
{\bf{o}}_n^{{\rm{out }}}(t) = \Big[ &{\left[ {i,{R_{i}}(t - 1),{\beta _{n,i}}(t - 1),} \right.} \nonumber\\
&\left. {\left. {{\beta _{n,i}}(t - 1)/{\beta _{i}}(t - 1)} \right]}_{i \in {\mathcal U}_n^{{\rm{out}}}(t - 1)} \right]. \nonumber
\end{align}


Similarly, given the set of the interfered AUs of BS $n$ is ${\mathcal U}^{\rm{pri}}$, we define the information from ${\mathcal U}^{\rm{pri}}$ as:
\begin{align}
{\bf{o}}_n^{{\rm{out}},{\rm{pri}}}(t) = \Big[& {\left[ {l,{\rho _l}(t - 1)/{I_{\max }},{\rho _{n,l}}(t - 1),} \right.} \nonumber\\
&{{{\left. {{\rho _{n,l}}(t - 1)/{\rho _l}(t - 1)} \right]}_{l \in {{\mathcal U}^{{\rm{pri}}}}}}} \Big], \nonumber
\end{align}
which can be obtained from the dedicated BSs serving the AUs in ${\mathcal U}^{\rm{pri}}$, such as ATG BS.

Therefore, in slot $t$, the observation of BS $n$ is:
\begin{align}
\label{State_Space}
{{\bf{o}}_n}(t) = \big[ &{\bf{o}}_n^{{\rm{loc }}}(t), {\bf{o}}_n^{{\rm{loc,pri}}}(t), {\bf{o}}_n^{{\rm{in}}}(t),\nonumber\\
&{{\bf{o}}^{{\rm{in,pri}}}}(t), {\bf{o}}_n^{{\rm{out }}}(t), {\bf{o}}_n^{{\rm{out,pri }}}(t) \big].
\end{align}


\addtolength{\topmargin}{-0.05in} 

\subsubsection{Action Space}
The beamforming vector can be calculated as ${{\bf{w}}_k} = \sqrt {{p_k}} {\overline {\bf{w}} _k}$.
By leveraging the structure of the solution derived from Algorithm \ref{alg_Closed_Form_FP_cog}, the parameters required to determine the beamforming vectors in massive cellular networks can be greatly reduced \cite{ge2024deep}.
Specifically, from the fifth line of Algorithm \ref{alg_Closed_Form_FP_cog}, we can learn that the normalized beamforming vector can be expressed as ${\overline {\bf{w}} _k} = \widetilde {\bf{D}}_{{\varrho _k}}^{ - 1}{{\bf{h}}_{{\varrho _k},k}}/{\left\| {\widetilde {\bf{D}}_{{\varrho _k}}^{ - 1}{{\bf{h}}_{{\varrho _k},k}}} \right\|_2}$, where ${\tilde{\bf{D}}}_n={\sum_{i = 1}^K {{\alpha _{n,i}}{{\bf{h}}_{n,i}}{\bf{h}}_{n,i}^H}  + \sum_{l = 1}^L {{\mu _{n,l}}{{\bf{g}}_{n,l}}{\bf{g}}_{n,l}^H}  + {\eta _n}{\bf{I}}}$.
It can be determined by the local CSI ${\bf{h}}_{{\varrho _k},i}, \forall i$ and ${\bf{g}}_{{\varrho _k},l}, \forall l$, some weights for the interference leakage power from BS $\varrho _k$ to TU $i$ and AU $l$, i.e., ${\alpha _{{\varrho _k},i}}$'s and ${\mu _{{\varrho _k},l}}$'s, and a scaling factor for noise $\eta _{\varrho _k}$.
%
%
Thus, we design the action space of BS $n$ as:
\begin{align}
\label{Action_Space}
{{\bf{a}}_n} = [& q_n^{{\rm{total}}},{q_{n,1}}, \cdots ,{q_{n,K}},{\alpha _{n,1}}, \cdots ,{\alpha _{n,K}},{\eta _n},\nonumber\\ 
&{\mu _{n,1}}, \cdots ,{\mu _{n,L}} ],
\end{align}
where $q_n^{{\rm{total }}} \in (0,1]$ represents the ratio of the total transmit power of BS $n$ to the maximum transmit power ${P_{\max }}$, 
${q_{n,k}} \in \left( {0,1} \right]$ represents the proportion of transmitted power that BS $n$ allocates to TU $k$, and $\sum_{k = 1}^K {{q_{n,k}}}  = 1$. 
Therefore, the power can be calculated as ${p_k} = {P_{\max }}q_n^{{\rm{total }}}{q_{n,k}}$.
The dimension of action space is $(2K+L+2)$.
Note that BS $n$ only calculates the beamforming vectors for its associated TUs, i.e., ${{\bf{w}}_k}, \forall k \in {{\cal K}_n}$, from its action ${{\bf{a}}_n}$.


\subsubsection{Reward Function}
According to the distributed reward function design in \cite{ge2024deep}, the reward function of BS $n$ is designed as:
\begin{equation}
\label{Reward_Function}
{r_n}\left( t \right) = \sum_{k \in {{\mathcal K}_n}\left( t \right)} {{R_k}} \left( t \right) - \sum_{i \in {\mathcal U}_n^{{\rm{out}}}\left( t \right)} {\left( {{R_{i\backslash n}}\left( t \right) - {R_i}\left( t \right)} \right)},
\end{equation}
where ${R_k}\left( t \right) = {\log _2}\left( {1 + p_k^r\left( t \right)/{\beta _k}\left( t \right)} \right)$ and ${R_{i\backslash n}}\left( t \right) = {\log _2}\left( {1 + p_i^{\rm{r}}\left( t \right)/\left( {{\beta _i}\left( t \right) - {\beta _{n,i}}\left( t \right)} \right)} \right)$. 
The second term in \eqref{Reward_Function} is a penalty for interfering with the interfered TUs of BS $n$.

\subsubsection{Cost Function}
The cost function of BS $n$ is ${{\bf{c}}_n}(t) = {\left[ {{c_{n,1}}(t), \cdots ,{c_{n,L}}(t)} \right]^T}$, which should reflects the received interference power of $L$ AUs in the constraint \eqref{cons_interfer}, i.e., ${\rho _l}(t) < {I_{\max }}, \forall l$.
We want the designed $l$-th cost function of BS $n$ ${c_{n,l}}(t)$ to be a straightforward linear function of ${\rho _l}(t)$, while ensuring that ${c_{n,l}}(t) < 0$ when ${\rho _l}(t) < {I_{\max }}$.
Accordingly, the cost limit $b_{n,l}$, $\forall l$ should be set to 0 and ${c_{n,l}}(t)$ can be designed as:
\begin{equation}
\label{Cost_Function}
{c_{n,l}}(t) = {\rho _l}(t)/{I_{\max }} - 1.
%
\end{equation}
Therefore, according to the NCPOMG formulation in \eqref{NCPOMG}, the safety constraints of BS $n$ are given by:
\begin{equation}
\label{safety_constraints}
{{\mathbb{E}}_{{{\bf{s}} \left( 0 \right)},{{\bf{a}}_n\left( 0 \right)},...}}\left[ {\sum_{t = 0}^\infty  {\gamma _c^t\left( {\rho _l}(t)/{I_{\max }} - 1 \right)} } \right] \le 0, \forall l.
\end{equation}

\subsection{TU Agent for User Association}
The adopted DRL algorithm, observation space, action space, and reward function of the TU agents are detailed as follows.

\subsubsection{Algorithm}
We apply the D3QN algorithm in each TU agent, which combines the ideas of double DQN and dueling DQN and has been successfully applied to user agents in user association tasks \cite{van2016deep, wang2016dueling, zhao2019deep}.
The update of the double DQN decouples action selection and action evaluation, which can prevent overestimation problems that occur during DQN updates \cite{van2016deep}.
Specifically, the online network ${Q_{\boldsymbol{\omega }}}$ with parameters ${\boldsymbol{\omega }}$ is used to select the action, while the target network ${Q_{{\boldsymbol{\omega }}^{-}}}$ with parameters ${\boldsymbol{\omega }}^{-}$ is used to evaluate the action.
Every time the agent updates the networks, $B_b^{TU}$ experiences $\langle {{\bf{o}}_{j}^{TU}}, {\varrho_j}, {r_{j}^{TU}}, {{\bf{o}}_{j + 1}^{TU}} \rangle$, $j = 1, \ldots, B_b^{TU}$ are sampled to form a batch.
The loss function of double DQN is: 
\begin{equation}
\label{loss_D3QN}
{{\mathcal L}_Q}\left( {\boldsymbol{\omega }} \right) = \frac{1}{{2{B_b^{TU}}}}\sum_{j = 1}^{{B_b^{TU}}} {{{\left( {{Q_{\boldsymbol{\omega }}}\left( {{{\bf{o}}_j^{TU}},{\varrho_j}} \right) - Q_j^{{\rm{target}}}} \right)}^2}},
\end{equation}
where $Q_j^{\rm{target}} \!=\! {r_j^{TU}} \!+ \gamma Q_{{\boldsymbol{\omega}}^{-} }\!\!\left( {{\bf{o}}_{j + 1}^{TU}}, {\arg\max }_{\varrho^{\prime}} Q_{\boldsymbol{\omega}}\!\left( {{\bf{o}}_{j + 1}^{TU}},{\varrho^{\prime}} \right) \!\right)$.
Besides, in dueling DQN, the output Q value of each action is combined from two streams that estimate the scalar state value and the advantage of each action \cite{wang2016dueling}.

\subsubsection{Observation Space}
The observation design is based on an intuitive understanding that if the rate obtained by TU $i$ associating with a BS is low, then TU $i$ can associate with another BS with fewer associated users and a stronger channel.
Thus, in slot $t$, the observation of TU $k$ is designed as:
\begin{align}
\label{association_observation}
{\bf{o}}_k^{TU}\!\left( t \right) =\! [& \left| {{{\mathcal K}_1}\left( {t -\! 1} \right)} \right|\!, \cdots\! ,\left| {{{\mathcal K}_N}\left( {t -\! 1} \right)} \right|\!,{\boldsymbol{\chi }}_k^{TU}(t -\! 1),{{\tilde{\bf{h}}} _k}\left( t \right)\!,\nonumber\\
&{p_k}\left( {t -\! 1} \right)\!,{R_k}\left( {t -\! 1} \right)\!,p_k^r\left( {t -\! 1} \right)\!,{\beta _k}\left( {t -\! 1} \right) ],
\end{align}
where $\left| {{{\mathcal K}_n}} \right|$ is the number of users associated with BS $n$, ${\boldsymbol{\chi }}_k^{TU} = \left[ {{\chi _{1,k}}, \cdots ,{\chi _{N,k}}} \right]$ is the user association for TU $k$, ${{\tilde{\bf{h}}} _k} = \left[ {\left\| {{{\bf{h}}_{1,k}}} \right\|_2^2, \cdots ,\left\| {{{\bf{h}}_{N,k}}} \right\|_2^2} \right]$ is the channel strength from all BSs to TU $k$.
At the beginning of each time step $t$, each BS broadcasts reference signals to all TUs, and then each TU can measure the channel strength from all BSs.
Besides, TU $k$ can obtain $\left| {{{\mathcal K}_n}\left( {t - 1} \right)} \right|, \forall n$ from BS $\varrho_k\left( {t - 1} \right)$.

\subsubsection{Action Space}
The action of TU $k$ is its associated BS, i.e., $\varrho_k$.

\subsubsection{Reward Function}
The reward function of TU $k$ is designed as:
\begin{align}
\label{Reward_Function_asso}
{r_k^{TU}}\left( t \right) =& {\widetilde R_k}\left( t \right) - \sum_{i \in {\mathcal U}_{{\varrho _k}\left( t \right)}^{{\rm{out }}}\left( t \right),i \ne k} {\left( {{{\widetilde R}_{i\backslash k}}\left( t \right) - {R_i}\left( t \right)} \right)},\\
\label{Reward_Function_rate_discount}
{\widetilde R_k}\left( t \right) =& \left\{ \begin{array}{l}
{R_k}\left( t \right),{\varrho_k}\left( t \right) = {\varrho_k}\left( {t - 1} \right),\\
{\zeta _R}{R_k}\left( t \right),{\varrho_k}\left( t \right) \ne {\varrho_k}\left( {t - 1} \right),
\end{array} \right.
\end{align}
where ${\zeta _R} \in \left[ {0,1} \right]$ is a handover discount factor, representing the fraction of time for data transmission in the time slot during which handover occurs, ${{\widetilde R}_{i\backslash k}}\left( t \right) - {R_i}\left( t \right)$ is a penalty term for the rate loss caused by the BS $\varrho_k$ to TU $i$ in order to serve TU $k$, and ${\widetilde R_{i\backslash k}}\left( t \right) = {\log _2}\left( {1 + {p_i^{\rm{r}}\left( t \right)}/ \left({{\beta _i}\left( t \right) - {p_k}\left( t \right){{\left| {{\bf{h}}_{{\varrho _k},i}^H\left( t \right){{\overline {\bf{w}} }_k}\left( t \right)} \right|}^2}}\right)} \right)$.

\subsection{The Overall Safe DRL Scheme}

The proposed safe DRL-based DDUACBF scheme for CATN is summarized in Algorithm \ref{alg_safeMADRL_UA_CBF}.
In each time slot, all TU agents first determine their associated BS, and then all BS agents determine their downlink beamforming vectors to serve their associated TUs.




\begin{algorithm}[t] 
\small
\caption{Safe DRL-based DDUACBF scheme for CATN}
\label{alg_safeMADRL_UA_CBF}
\begin{algorithmic}[1]
\STATE {TU $k$ initializes $Q_{\boldsymbol{\omega}}$ and $Q_{\boldsymbol{\omega}^{-}}$ by ${\boldsymbol{\omega}^{-}} \leftarrow {\boldsymbol{\omega}}$, $\forall k$.}
\STATE {TU $k$ initializes discount factor $\gamma$, learning rates $\alpha_{\boldsymbol{\omega}}$, $\epsilon$, target D3QN replacement frequency $T^{-}$, and the experience replay memory $\xi _k^{TU}$ with a FIFO queue of size $B_m^{TU}$, $\forall k$.}
\STATE {BS $n$ initializes $\pi_{\boldsymbol{\vartheta}}$, $V_{\boldsymbol{\varphi}}$, and $V^C_{\boldsymbol{\psi}}$, $\forall n$.}
\STATE {BS $n$ initializes discount factor $\gamma$, cost limit ${\bf{b}}$, GAE parameter $\lambda$, learning rates $\alpha_{\boldsymbol{\nu}}$, $\alpha_V$, $\alpha_{\pi}$, cost parameter ${\boldsymbol{\nu}}$ and its bound ${\boldsymbol{\nu}}_{\max}$, KL divergence bound $\varepsilon$, and the trajectory collector $\xi _n^{BS}$ of size $B_m^{BS}$, $\forall n$.}
\STATE {In time slot $t=0$, TU $k$ obtains a random action ${\varrho_k}(0)$, $\forall k$.}
\STATE {BS $n$ obtains the CSI to all TUs, the CSI to all AUs, and a random action ${\bf{a}}_n(0)$, $\forall n$. Set $t=1$.}
%
\WHILE {the training process is not over}
\STATE {When time slot $t$ begins, TU $k$ measures the channel strength from all BSs and obtains observation ${\bf{o}}_k^{TU}(t)$, $\forall k$.}
\STATE {If ${{\bf{o}}_k^{TU}}(t-1)$ exists, TU $k$ stores the experience $\left\langle {{{\bf{o}}_k^{TU}}(t-1),{\varrho_k}(t-1),{r_k^{TU}}\!(t-1),{{\bf{o}}_k^{TU}}(t)} \right\rangle$ into $\xi _k^{TU}$, $\forall k$.}
\IF {$t-1 \geq B_b^{TU}$}
\STATE TU $k$ samples a batch of $B_b^{TU}$ experiences $\langle {{\bf{o}}_{j}^{TU}}\!, {\varrho_j}, {r_{j}^{TU}}\!, {{\bf{o}}_{j + 1}^{TU}} \rangle$, $\!j \!= \!1, \ldots\!, B_b^{TU}\!$ from $\!\xi _k^{TU}\!$, $\!\forall k$.
\vspace{-0.3cm}
\STATE TU $k$ updates ${\boldsymbol{\omega}}  \leftarrow {\boldsymbol{\omega}}  - {\alpha _{\boldsymbol{\omega}}}{\nabla _{\boldsymbol{\omega}} }{{\mathcal L}_Q}\left( {\boldsymbol{\omega }} \right)$ with \eqref{loss_D3QN}, $\forall k$.
\STATE In every $T^{-}$ slots, TU $k$ replaces the parameters of the target D3QN ${\boldsymbol{\omega}^{-}} \leftarrow {\boldsymbol{\omega}}$, $\forall k$.
\ENDIF
\IF {$t-1 \geq B_b^{TU}$}
\STATE {TU $k$ obtains action ${\varrho_k}(t)$ through $\epsilon$-greedy policy and associates with BS ${\varrho_k}(t)$, $\forall k$.}
\ELSE
\STATE {TU $k$ obtains a random action ${\varrho_k}(t)$, $\forall k$.}
\ENDIF
\STATE {BS $n$ obtains the CSI to all TUs and the CSI to all AUs, $\forall n$.}
\STATE {BS $n$ exchanges information with other BSs, $\forall n$.}
\STATE {BS $n$ obtains observation ${\bf{o}}_n(t)$, and stores the experience $\left\langle {{{\bf{o}}_n}(t-1),{{\bf{a}}_n}(t-1),{r_n}(t-1),{{\bf{c}}_n}(t-1), {{\bf{o}}_n}(t)} \right\rangle$ into $\xi _n^{BS}$ if ${{\bf{o}}_n}(t-1)$ exists, $\forall n$.}
\IF {if $\xi _n^{BS}$, $\forall n$ is full}
\STATE BS $n$ obtains all experiences $\langle {{{\bf{o}}_{i}},{{\bf{a}}_{i}},{r_{i}},{{\bf{o}}_{i + 1}},{{\bf{c}}_{i}}} \rangle$, $i = 1, \!\ldots\!, B_m^{BS}\!$ in $\xi _n^{BS}\!$ and updates $\pi_{\boldsymbol{\vartheta}}$, $\!V_{\boldsymbol{\varphi}}$, and $V^C_{\boldsymbol{\psi}}\!$ by the safe DRL algorithm, i.e., Algorithm \ref{alg_CUP}, and then clear $\xi _n^{BS}$, $\forall n$.
\ENDIF
\STATE {BS $n$ obtains action ${\bf{a}}_n(t)$ and calculates ${{\bf{w}}_k}, \forall k \in {{\cal K}_n}$, $\forall n$.}
\STATE {BS $n$ transmits signals to its associated TUs, then receives reward $r_n(t)$ and cost ${{\bf{c}}_n}(t)$, $\forall n$.}
\STATE {TU $k$ receives reward $r_k^{TU}(t)$, $\forall k$. Set $t=t+1$.}
\ENDWHILE
\RETURN $\!\!\!$ Policy network $\!\pi_{\boldsymbol{\vartheta}}\!$ of each BS and D3QN $\!Q_{\boldsymbol{\omega}}\!$ of each TU.
\end{algorithmic}
\end{algorithm}


\section{Simulation Results}
\label{sec_Simulation_Results}
In this section, simulation results are shown to evaluate the performance of the proposed safe DRL-based DDUACBF scheme in Algorithm \ref{alg_safeMADRL_UA_CBF} (denoted by D3QN-CUP) and some two-stage benchmarks, which first determine the UA and then the beamforming vectors in each time slot.
For UA, the benchmarks are as follows:
\begin{itemize}
\item DCD: The UA in each time slot is obtained by the joint UA and power control algorithm in \cite{shen2014distributed}, which iteratively optimizes the UA and the power control via the DCD algorithm in Algorithm \ref{alg_DCD} and the Newton's method, respectively. This algorithm requires real-time global channel strength information.
\item SC: Each TU is associated with the BS with the strongest channel strength. Thus, each TU only needs the received channel strength information from all BSs without information exchange.
\end{itemize}
For beamforming, the benchmarks are as follows:
\begin{itemize}
\item DFP: The beamformers are obtained by the direct FP algorithm with real-time global CSI, which can achieve a higher sum rate than the WMMSE algorithm and provides upper bound performance \cite{shen2018fractional1}.
\item PPO: The beamformers are obtained by PPO algorithm \cite{schulman2017proximal}, where the reward function is designed as ${\widetilde r_n}(t) = {r_n}(t) - \zeta \sum_{l = 1}^L {\max \left\{ {{c_{n,l}}(t),0} \right\}}$ with \eqref{Reward_Function}, \eqref{Cost_Function}, and a fixed penalty weight $\zeta$. 
Besides, the observation space and the action space are given by \eqref{State_Space} and \eqref{Action_Space}, respectively.
\item WMMSE: The beamformers are obtained by a low-complexity modified version of Algorithm \ref{alg_Closed_Form_FP_cog} with real-time global CSI.
Specifically, the high-complexity matrix inversion operation in line 5 is replaced by updating ${{\bf{w}}_k}$'s and $\eta _n$'s by the prox-linear method \cite{xu2013block} and updating ${\mu _l}$'s by the sub-gradient method.
\end{itemize}

\begin{figure}[t]
\centering
\includegraphics[width=9cm]{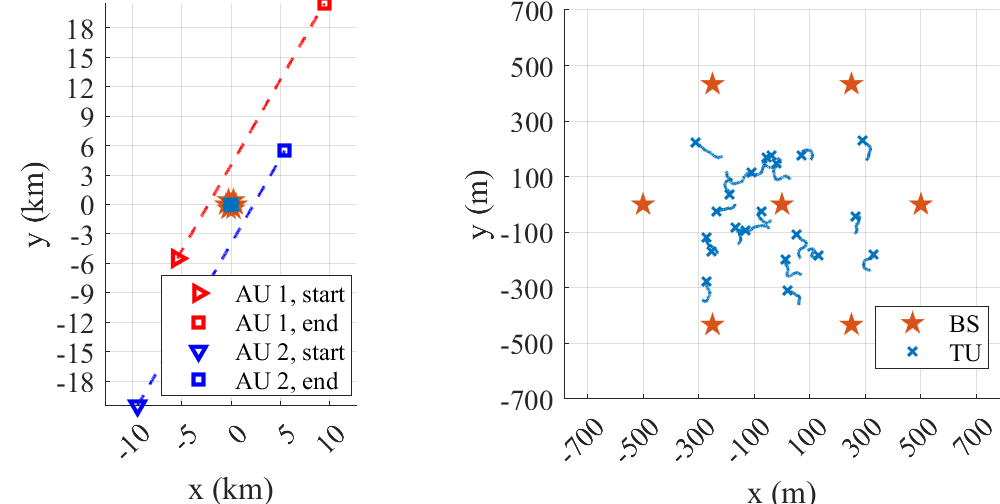}
\caption{Simulation scenario.}
\label{simulation_scenario}
\vspace{-0.4cm}
\end{figure}
We consider a CATN with $L=2$ AUs, $N=7$ BSs, and $K=21$ TUs.
The projection positions of each device on the ground are shown in Fig. \ref{simulation_scenario}, where the AUs are airplanes with a fixed height of $10{\rm{km}}$.
Note that the right figure in Fig. \ref{simulation_scenario} is an enlarged view of the middle part of the left figure.
Besides, the heights of the terrestrial BSs and TUs are $30{\rm{m}}$ and $1.5{\rm{m}}$, respectively.
We consider 6000 time slots with an interval of $0.02{\rm{s}}$, where both AUs fly with a fixed speed of $250{\rm{m/s}}$.
The movement trajectories of TUs are depicted by the blue curve in Fig. \ref{simulation_scenario}.
%
%
Other parameters are set as follows.
The size of the URA at each BS is set as $M_h = M_v = 4$, i.e., $M = 16$.
The center frequency is $f_c=2{\rm{GHz}}$ and the bandwidth is $10{\rm{MHz}}$.
The noise power is set to $-114{\rm{dBm/MHz}}$, i.e., $\sigma^2=3.98\times 10^{-11}{\rm{mW}}$.
For each channel ${{\bf{h}}_{n,k}}$, its path loss $L_{{\rm{UMa}},n,k}$ keeps LoS or NLoS during the simulation.
The channel correlation coefficient is $\alpha=0.64$.
The Rician factor is $\kappa=15{\rm{dB}}$.
The handover discount factor ${\zeta _R}$ is 0.4.

The size of the codebook is set as $C = 128$ and the compression factor is set as $N_c = 4$. 
For the size of the sets, we set $\left| {{\mathcal B}_{n,k}^{{\rm{in}}}(t)} \right| = B_{\rm{in}}=4$, $\left| {{\mathcal B}_l^{{\rm{in,pri}}}(t)} \right| = \widetilde B_{\rm{in}}=5$, and $K_{\rm{out}}=3$, then we have $\left| {{\mathcal U}_n^{{\rm{out }}}(t)} \right| = K_{\rm{out}} B_{\rm{in}}=12$.
%
For the BS agent in the CUP-based schemes and the PPO-based schemes, the policy and value networks' hidden neural networks are all three-layer with 512, 128, and 64 neurons, respectively. 
Besides, in the value network and the first two hidden layers of the policy network, ReLU is adopted as the activation function. 
In the third hidden layer of the policy network, the sigmoid activation function is adopted.
Moreover, the sigmoid function is also adopted to normalize the outputs of the policy network to the range $[0, 1]$ before scaling to the actual action values.
For the TU agent in the D3QN-based schemes, the Q networks' hidden neural networks are all two-layer with 64 and 32 neurons. Besides, ReLU is adopted as the activation function.
The parameters of Algorithm \ref{alg_safeMADRL_UA_CBF} are set as follows: 
discount factor is $\gamma=0.5$, 
GAE parameter is $\lambda=0.1$,
learning rates are $\alpha_{\boldsymbol{\nu}}=\!0.06$, $\alpha_V\!=\!3\times 10^{-4}$, $\alpha_{\pi}\!=\!3\times 10^{-4}$, 
each element of initial cost constraint parameter ${\boldsymbol{\nu}}$ is set as 1, 
each element of cost constraint parameter bound ${\boldsymbol{\nu}}_{\max}$ is set as 10, 
cost bound is ${\bf{b}}=\!{\bf{0}}$, 
KL divergence bound is $\varepsilon=\!0.02$. 
Besides, for BS agent, we set $B_m^{BS}=50$, minibatch size $B_b=10$, and number of epoch $B_e=20$.
For TU agent, we set $T^{-}=50$, $B_m^{TU}=2000$, and batch size $B_b^{TU}=200$.
Each time the TU agent obtains action through $\epsilon$-greedy policy, its $\epsilon$ is updated by $\epsilon = \max \left\{ {{\epsilon_{\min }},0.995\epsilon} \right\}$, where $\epsilon_{\min }=0.005$ and $\epsilon$ is initialized to 0.3.



\begin{figure}[t]
\centering
\subfloat[Sum rate of the terrestrial users. \label{sum_rate_time_slots}]
{\includegraphics[width=6.5cm]{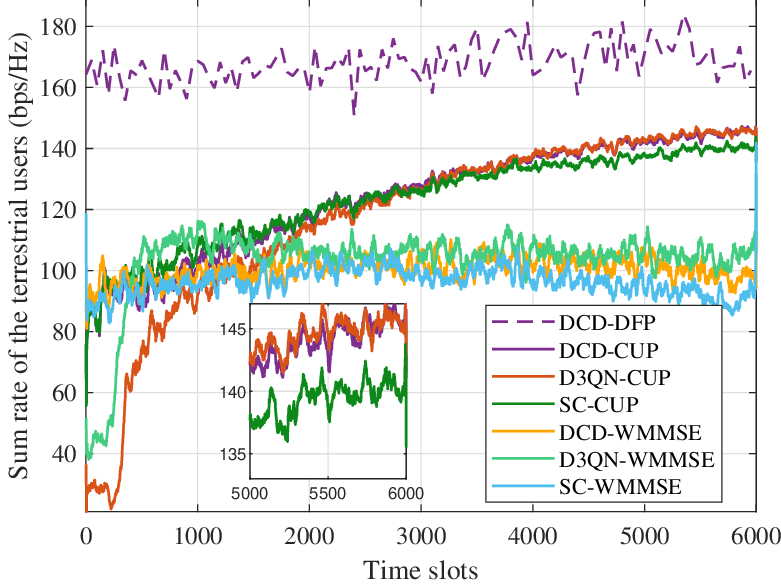}}
\vspace{-0.2cm}
\\
\subfloat[Received interference power of two aerial users. \label{interference_power_time_slots_PUs}]
{\includegraphics[width=6.5cm]{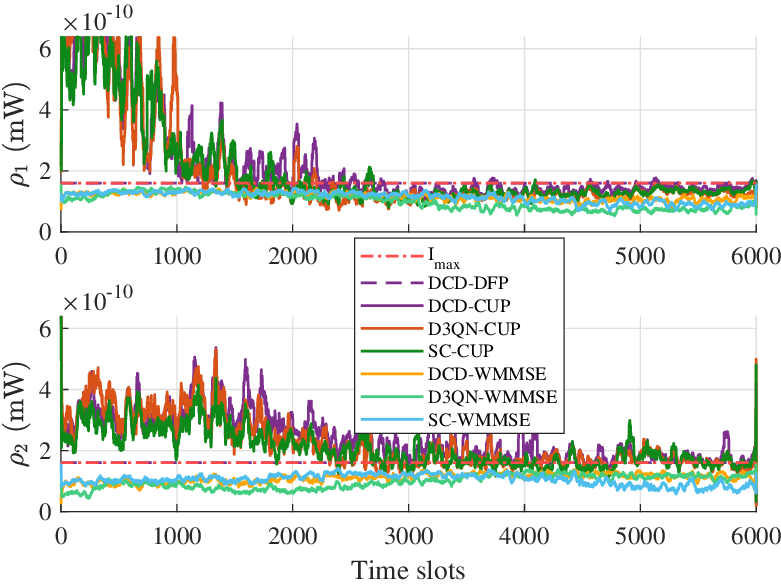}}
%
\caption{Performance comparison of the proposed scheme and the benchmarks: $P_{\max}=20{\rm{W}}$ and ${I_{\max}} = 1.6\times {10^{ - 10}}{\rm{mW}}$.}
\label{time_slots_CUP_WMMSE} 
\vspace{-0.3cm}
\end{figure}

First, we show the convergence performance of the proposed D3QN-CUP scheme and the benchmarks.
Due to the high computational complexity of the DFP-based scheme, we choose 100 time slots of equal time intervals to compute its results, which are plotted by dashed lines.
In addition, in all simulation figures, each value is a moving average with a span of 41 time slots.
Fig. \ref{time_slots_CUP_WMMSE}(a) shows the sum rate of the TUs during training.
The sum rate obtained by the CUP-based schemes gradually increases and is higher than that of the corresponding WMMSE-based schemes after convergence. 
Besides, the D3QN-based schemes can achieve a higher sum rate compared to the corresponding SC-based schemes.
Accordingly, Fig. \ref{time_slots_CUP_WMMSE}(b) shows the received interference power of two AUs ${\rho _{l}}, \forall l$ during training. 
The power ${\rho _{l}}, \forall l$ obtained by the CUP-based schemes gradually reduces and is generally close to $I_{\max}$ after convergence.



\begin{figure}[t]
\centering
\subfloat[Sum rate of the terrestrial users. \label{sum_rate_time_slots_penalty_factor}]
{\includegraphics[width=6.5cm]{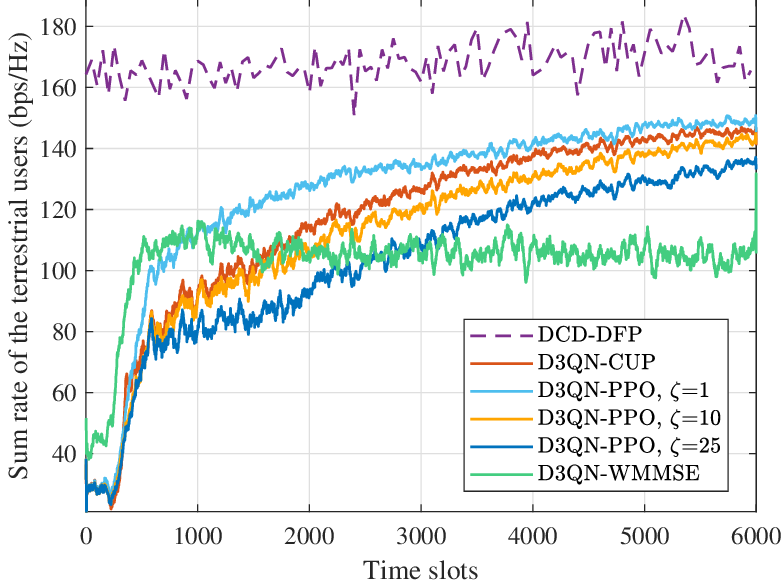}}
\vspace{-0.2cm}
\\
\subfloat[Received interference power of two aerial users. \label{interference_power_time_slots_PUs_penalty_factor}]
{\includegraphics[width=6.5cm]{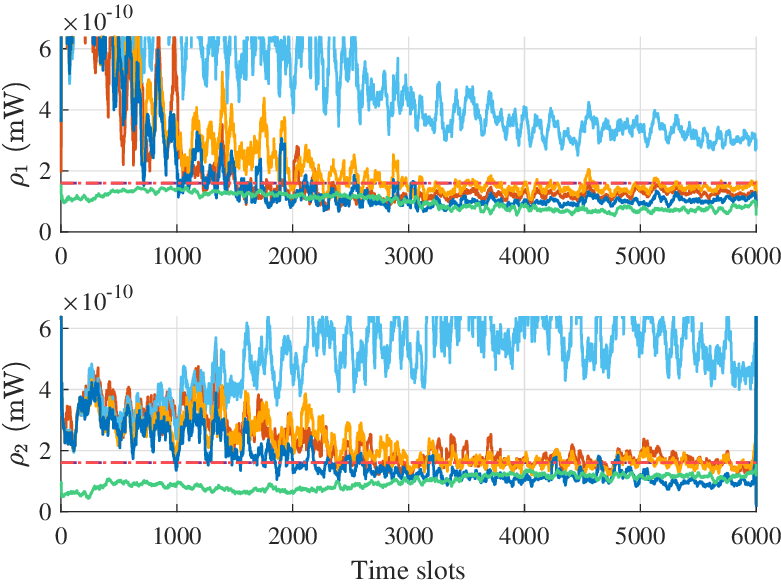}}
%
\caption{Performance of the penalty-based PPO with different penalty weights $\zeta$: $P_{\max}=20{\rm{W}}$ and ${I_{\max}} = 1.6\times {10^{ - 10}}{\rm{mW}}$.}
\label{time_slots_penalty_factor} 
\vspace{-0.3cm}
\end{figure}
    
Fig. \ref{time_slots_penalty_factor} shows the performance of the D3QN-PPO scheme with different penalty weights $\zeta$.
When $\zeta$ is large, the D3QN-PPO scheme is too conservative to achieve a high sum rate and takes a long time to converge in Fig. \ref{time_slots_penalty_factor}(a). 
When $\zeta$ is small, the D3QN-PPO scheme is too bold to effectively guarantee the constraint satisfaction in Fig. \ref{time_slots_penalty_factor}(b), despite achieving a higher sum rate than the D3QN-CUP scheme. 
Thus, the performance of the D3QN-PPO scheme is sensitive to the penalty weight $\zeta$.
In contrast, without tricky penalty weight adjustments, the D3QN-CUP scheme can maximize the reward while satisfying the safety constraints, which avoids the economic expenses of multiple training attempts in real-world deployments.

\begin{figure}[t]
\centering
\subfloat[Sum rate of the terrestrial users. \label{sum_rate_time_slots_max_interference}]
{\includegraphics[width=6.5cm]{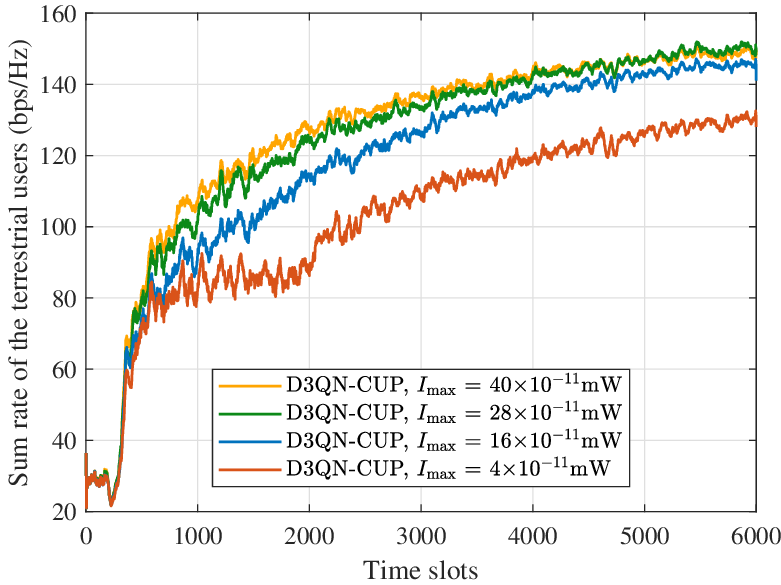}}
\vspace{-0.2cm}
\\
\subfloat[Received interference power of two aerial users. \label{interference_power_time_slots_PUs_max_interference}]
{\includegraphics[width=6.5cm]{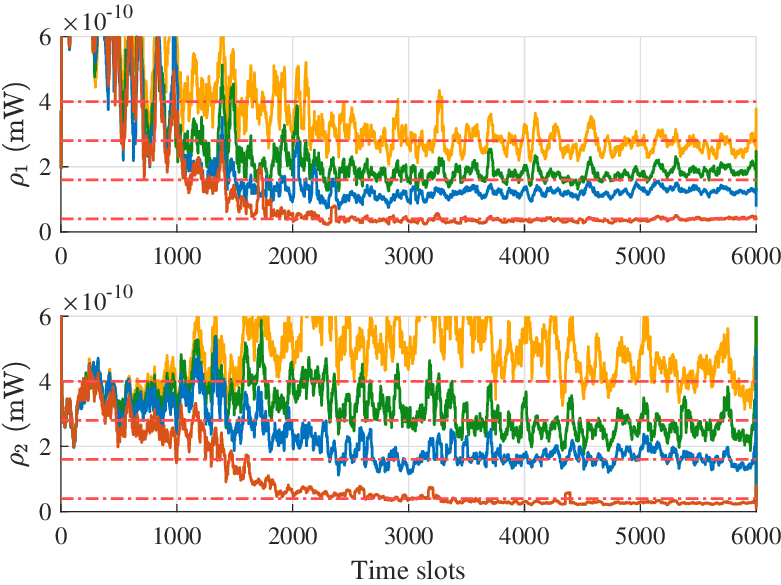}}
\caption{Performance for different interference temperature limits ${I_{\max}}$: $P_{\max}=20{\rm{W}}$.}
\label{time_slots_max_interference} 
\vspace{-0.3cm}
\end{figure}
Fig. \ref{time_slots_max_interference} shows the performance of the proposed D3QN-CUP scheme with different interference temperature limits ${I_{\max}}$.
In the case of different ${I_{\max}}$, the proposed scheme can gradually maximize the sum rate of the terrestrial network and reduce the interference power to the two AUs to near the threshold ${I_{\max}}$.
In addition, as ${I_{\max}}$ decreases, the sum rate also decreases to meet the more stringent interference requirements.

\begin{figure}[t]
\centering
\subfloat[Sum rate of the terrestrial users. \label{sum_rate_time_slots_max_power}]
{\includegraphics[width=6.5cm]{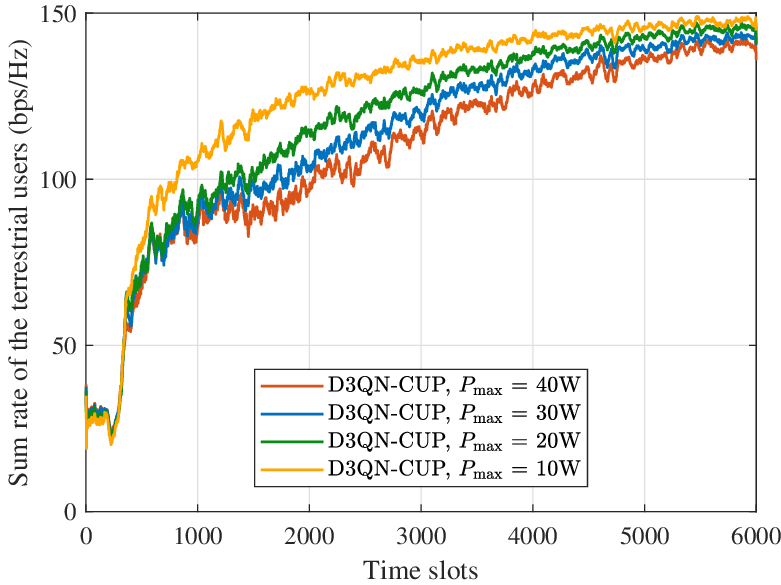}}
\vspace{-0.2cm}
\\
\subfloat[Received interference power of two aerial users. \label{interference_power_time_slots_PUs_max_power}]
{\includegraphics[width=6.5cm]{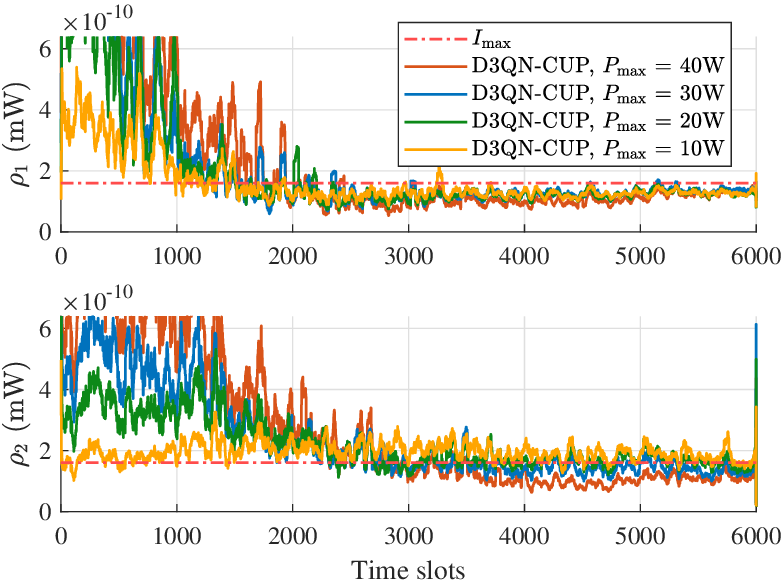}}
\caption{Performance for different maximum transmit power of the BSs ${P_{\max}}$: ${I_{\max}} = 1.6\times {10^{ - 10}}{\rm{mW}}$.}
\label{time_slots_max_power} 
\vspace{-0.3cm}
\end{figure}
Fig. \ref{time_slots_max_power} shows the performance of the proposed D3QN-CUP scheme with different maximum transmit power of the BSs ${P_{\max}}$.
It can be observed that as ${P_{\max}}$ increases, the proposed scheme requires more training slots to converge.
This may be attributed to the expanded action range resulting from the higher ${P_{\max}}$, which makes the policy need more exploration to reach the optimum.
Another possible reason is that when ${P_{\max}}$ is large, for fear that other agents will interfere too much with the AUs, the agents tend to adopt a more conservative policy to satisfy the safety constraints.

%
%
\begin{figure}[t]
\centering
\includegraphics[width=8.5cm]{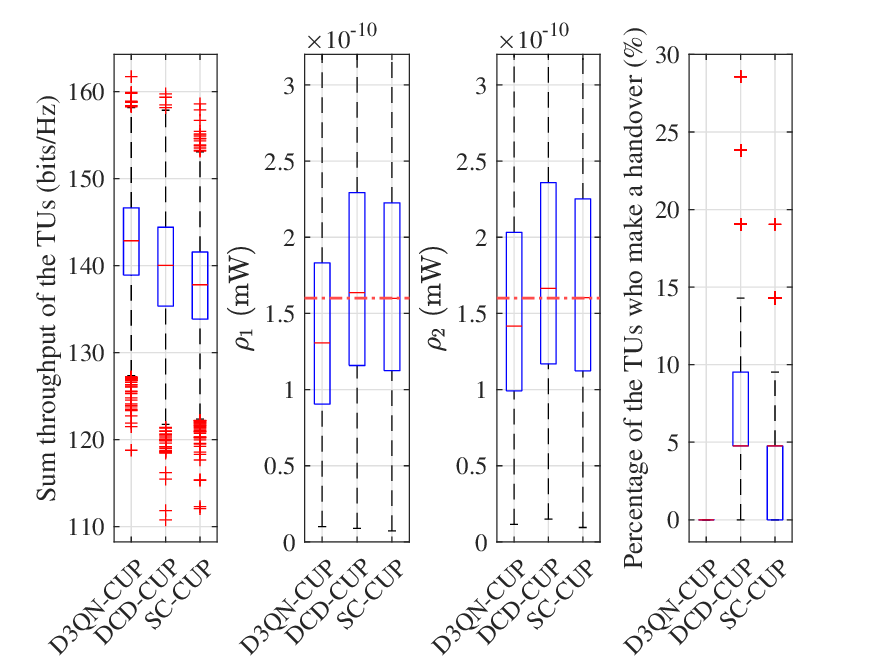}
\caption{Performance for different user association schemes: $P_{\max}=20{\rm{W}}$ and ${I_{\max}} = 1.6\times {10^{ - 10}}{\rm{mW}}$.}
\label{simulation_user_association}
\vspace{-0.4cm}
\end{figure}
Fig. \ref{simulation_user_association} shows the sum throughput of the TUs in one second, the received interference power of the two AUs, and the percentage of TUs that perform a handover obtained by the proposed D3QN-CUP scheme, the DCD-CUP scheme, and the SC-CUP scheme in 6000 test time slots, respectively.
It can be found that the D3QN-CUP scheme achieves the highest throughput, the lowest received interference power, and the lowest handover percentage.

\begin{table*}[t]
\centering  %
\caption{Comparison of communication overhead and computational complexity}  %
\label{table_Comparison}  %
\begin{tabular}{cccc}  
\toprule  %
Scheme  & \makecell[c]{External information required\\ (for user association)} & \makecell[c]{Inter-BS communication overhead\\ (for beamforming)} & \makecell[c]{Average computation time\\($L=2$, $N=7$, $K=21$, $M=16$)} \\  
\midrule  %
DCD-DFP & $N(K-1)$ & $2M(N-1)K + 2M(N-1)L$ & 78.78s \\ 
DCD-CUP & $N(K-1)$ & \multirow{4}*{\makecell[c]{$\left( {3{N_c}K + K + 2} \right){B_{{\rm{in}}}}{K_{{\rm{in}}}} +$\\ $\left( {K + 6} \right){\tilde B_{{\rm{in}}}}L + 3{K_{{\rm{out}}}}{B_{{\rm{in}}}} + 2L$}} & 326.80ms \\ 
D3QN-CUP (proposed) & $N$ & ~ & 82.64ms \\ 
SC-CUP & 0 & ~ & 63.48ms \\ 
D3QN-PPO & $N$ & ~ & 78.72ms \\ 
D3QN-WMMSE  & $N$ & $2M(N-1)K + 2M(N-1)L$ & 1.257s \\ 
\bottomrule  %
\end{tabular}
\vspace{-0.4cm}
\end{table*}
In Table \ref{table_Comparison}, we compare the communication overhead and the computational complexity of the proposed scheme and the benchmarks.
For clarity, we categorize the communication overhead into two parts: user association and beamforming. 
The external information required for user association is measured by the number of real scalar values that each TU needs to obtain from BSs.
The inter-BS communication overhead is measured by the number of real scalar values that each BS needs to obtain from other BSs.
Note that the information obtained by the CUP or PPO-based schemes through exchange is all from the previous time slot.
It can be found that when the number of antennas $M$ is large, the CUP or PPO-based schemes are more advantageous than the DFP or WMMSE-based schemes.
In our simulation setup, the inter-BS communication overhead required for the CUP or PPO-based schemes is 3610, while that for the DFP or WMMSE-based schemes is 4416.

Moreover, the computational complexity is measured by the average computation time to obtain the user association of all TUs and the beamforming vectors of all BSs of one time slot in the considered simulation scenario.
We run programs of these schemes on a computer equipped with Intel i7-11700K and NVIDIA GeForce RTX 2080 Ti.
Note that the DCD-DFP scheme is run in MATLAB, while other schemes are programmed in Python.
It can be seen that for the computational complexity of user association, the D3QN-based schemes are slightly higher than the SC-based schemes and significantly lower than the DCD-based schemes.
For beamforming, the computational complexity of the CUP or PPO-based schemes is considerably lower than that of the DFP or WMMSE-based schemes.
Thus, the proposed scheme demonstrates lower computational complexity than the iterative optimization-based schemes.
As illustrated in Table \ref{table_Comparison}, the proposed scheme with low communication overhead and computational complexity is more advantageous in the dynamic CATN environment.

\section{Conclusions}
\label{sec_Conclusions}
In this paper, we have proposed a safe DRL-based DDUACBF scheme for the CATN.
Specifically, to maximize the sum rate of the TUs under the interference temperature constraints of the AUs, a problem has been formulated by optimizing the user association between TUs and BSs and the beamforming vectors of the terrestrial BSs.
Then, this problem has been modeled as an NCPOMG with cost functions derived from the received interference power of the AUs.
To solve the NCPOMG, a safe DRL-based DDUACBF scheme has been proposed, where a safe DRL algorithm is adopted in each BS agent to maximize reward while satisfying safety constraints.
Simulation results have shown that the proposed scheme has lower computational complexity and communication overhead than the optimization-based schemes.
Moreover, the proposed scheme can achieve a high sum rate of the TUs while the average received interference power of the AUs is generally below the threshold.


\small
\bibliographystyle{IEEEtran}

\bibliography{IEEEabrv,CogMultiCellBF}
\end{document}